\documentclass[conference]{IEEEtran}

\usepackage{cite}
\usepackage{amsmath,amssymb,amsfonts}
\usepackage{algorithm}
\usepackage{algorithmic}
\usepackage{graphicx}
\usepackage{textcomp}
\usepackage{xcolor}
\usepackage[utf8]{inputenc}
\usepackage{subcaption} 
\usepackage{booktabs}  
\graphicspath{ {./} } 
\usepackage{url}
\usepackage{tabularx}
\usepackage{comment}

\def\BibTeX{{\rm B\kern-.05em{\sc i\kern-.025em b}\kern-.08em
T\kern-.1667em\lower.7ex\hbox{E}\kern-.125emX}}

\begin{document}

\title{GDEV-AI: A Generalized Evaluation of Deep Learning Inference Scaling and Architectural Saturation}

\author{\IEEEauthorblockN{Kathiravan Palaniappan}
\IEEEauthorblockA{\textit{Independent Researcher (University of Colorado Colorado Springs Alumni)} \\
kpalania@uccs.edu}
}

\maketitle

\begin{abstract}
The deployment of deep learning inference in production environments is on the rise, where factors like 
throughput, latency, and hardware efficiency are crucial. Despite a shift towards specialized hardware, 
numerous inference tasks still rely on CPU-only systems, particularly in legacy data centers and 
cost-sensitive environments. This study explores the practical scalability constraints of CPU-based 
inference for convolutional neural networks by benchmarking ResNet~\cite{he2016resnet} models 
across various batch sizes on a dual-tier hardware configuration: a legacy \textbf{Intel Xeon E5-2403 v2} 
and a modern \textbf{Intel Xeon 6 "Granite Rapids" (GNR)} processor. 

Our findings reveal that the inference throughput of legacy CPUs rapidly reaches saturation, often struggling to 
scale beyond a small batch size due to limitations in instruction sets and memory bandwidth. This early saturation 
underscores a fundamental architectural mismatch~\cite{ben2019benchmarking} 
between modern tensor-heavy workloads and traditional general-purpose processors. In contrast, the Granite Rapids system leverages 
\textbf{Intel Advanced Matrix Extensions (AMX)} to achieve substantial improvements in throughput. However, we also identify 
a critical "performance cliff" in the modern architecture; by testing thread counts beyond the physical core limit (oversubscription), 
we demonstrate how context-switching and execution unit contention lead to substantial tail-latency degradation, while 
latency increases linearly as computational and memory constraints 
are surpassed. To address the need for 
systematic performance analysis, we present 
\textbf{GDEV-AI (Generalized Deep Learning Evaluation for AI)}, a reproducible benchmarking framework designed 
to assess scalability limits and architectural saturation. By establishing a vendor-neutral 
CPU-only baseline, GDEV-AI provides the empirical evidence necessary to identify performance 
bottlenecks and guide capacity planning in heterogeneous data centers. This paper lays the groundwork for the broader 
GDEV-AI research initiative, which will systematically evaluate inference performance across evolving hardware tiers through a series of focused studies.
\end{abstract}


\section{Introduction}
Throughout their evolution, CPUs have been designed to support a wide range of computational workloads, 
ranging from control-oriented logic to large-scale data processing. The growing integration 
of AI inference into general-purpose applications signifies a new phase in this evolution, where 
CPUs are now expected to handle machine learning workloads directly as part of regular application 
execution. This shift challenges traditional assumptions about general-purpose processor design and necessitates 
a re-evaluation of how CPUs are architected and provisioned to accommodate emerging AI-driven workloads. 
This trend suggests a broader architectural transition toward ubiquitous AI execution on general-purpose 
processors~\cite{hazelwood2018facebook}.

By evaluating performance on legacy tiers and comparing them with modern architectural advancements such as \textbf{Intel Advanced Matrix Extensions (AMX)} present in newer \textbf{Granite Rapids} processors, GDEV-AI provides a crucial decision-making framework for sustainability and cost-optimization in global data centers, where heterogeneous hardware remains the operational reality.

\begin{figure}[t]
  \centering
  \includegraphics[width=\linewidth]{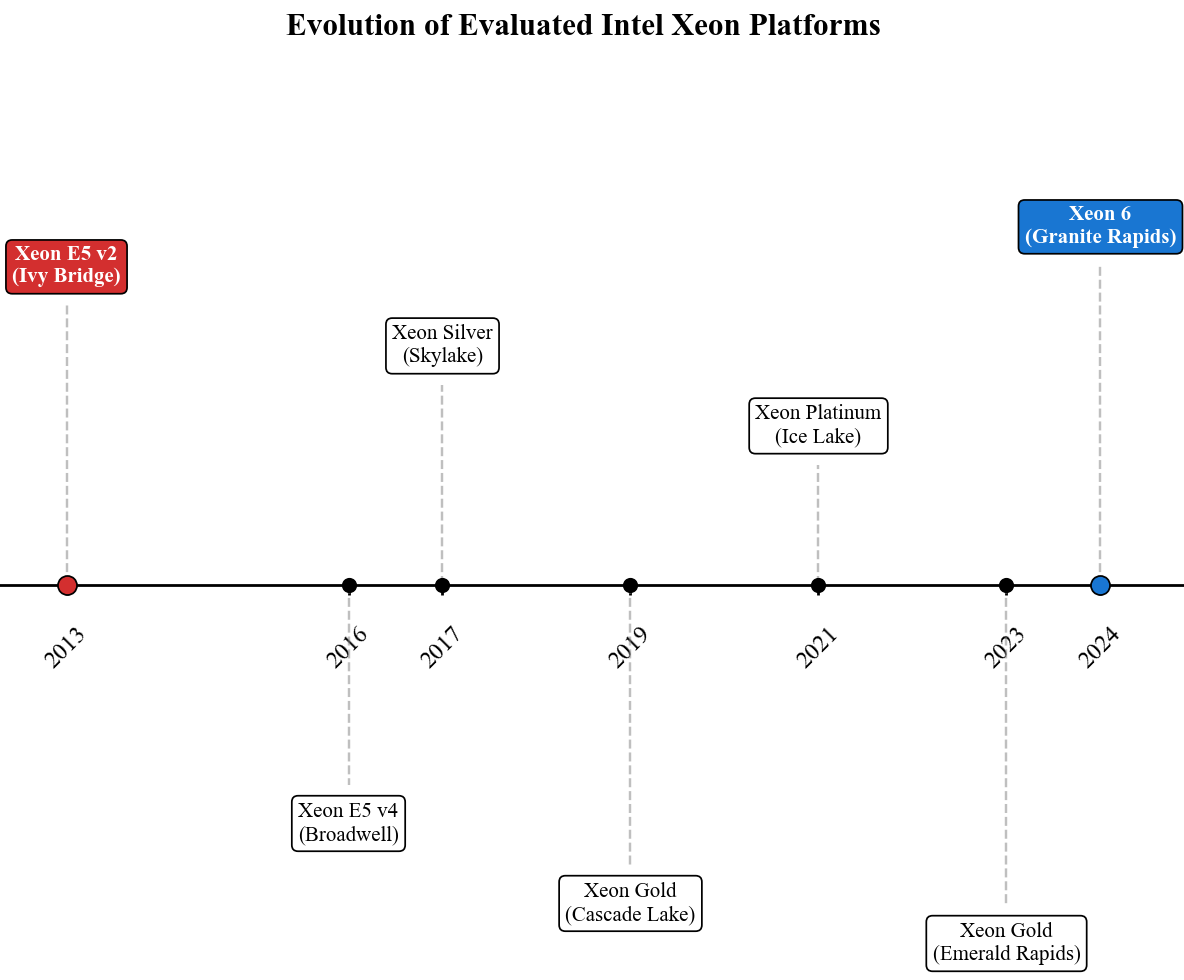}
  \caption{Evolution of Intel Xeon server processors across generations, 
  showcasing representative SKUs over time. This figure places the legacy 
  Xeon E5-2403 v2 platform, evaluated in this study, in context with a 
  modern Granite Rapids-class Xeon system. Both are experimentally assessed 
  to quantify generational performance scaling and architectural efficiency.}
  \label{fig:xeon_timeline}
\end{figure}

Deep learning has transitioned from research environments to large-scale production systems, supporting 
tasks ranging from image classification to recommendation engines. While a limited number of well-funded organizations 
oversee most model training, inference—the continuous execution of pre-trained models—is distributed 
across millions of production nodes \cite{hazelwood2018facebook}. Although the computational 
costs associated with training deep neural networks are well-documented, the cumulative resource consumption 
of inference has become a major contributor to modern data center workloads 
\cite{hazelwood2018facebook, mlperf_inference, crankshaw2017clipper}.

Initial model training requires significant computational resources and large datasets, but 
this expense is primarily a one-time occurrence. Once a model is mature, further training usually 
involves incremental updates with smaller datasets, leading to much lower utilization of training 
resources. In contrast, inference consumes the majority of computational resources over 
a model’s lifetime, as it is continuously executed to support real-world applications.

Despite the growing availability of specialized AI accelerators, many organizations still rely 
on traditional CPU-based systems for inference. This reliance is often due to practical constraints, 
such as legacy infrastructure, cost considerations, and limited access to GPUs. However, 
modern deep learning models like convolutional neural networks impose computational demands 
that differ significantly from traditional CPU workloads, requiring high levels of parallelism 
that general-purpose processors were not originally designed to provide.

Although new architectural advances continue to emerge, data centers do not quickly refresh their 
deployed systems. Legacy architectures typically remain in service for extended periods before 
being retired. During this time, these older platforms must support modern workloads. Understanding 
their performance limits is therefore crucial to ensure reliable operation and informed capacity 
planning.

Low-core-count server-class CPUs, similar to the platform evaluated in this study, remain widely 
deployed in heterogeneous production environments. Within these mixed-hardware clusters, such 
systems are often tasked with dedicated inference workloads alongside more modern high-density 
nodes, such as the \textbf{Xeon 6 (Granite Rapids)} architecture. In these deployments, legacy processors 
execute pre-trained models for production decision-making, 
making it essential to establish a performance baseline to understand how these older architectural 
tiers contribute to the overall latency and throughput of the global inference pipeline compared to modern 
counterparts featuring dedicated AI acceleration hardware.

CPUs are optimized for general-purpose execution and offer limited parallel execution resources compared 
to modern accelerators. As batch sizes increase, inference performance on CPUs is often constrained 
by memory bandwidth, cache hierarchy~\cite{williams2009roofline}, and limited vector execution 
capabilities~\cite{ben2019benchmarking}. For convolution-heavy models such as ResNet, these constraints 
can lead to early performance saturation, where additional batching no longer improves throughput and 
instead increases latency.

In contrast, GPUs are engineered to leverage extensive data parallelism, featuring thousands of lightweight 
cores and high-bandwidth memory systems optimized for tensor operations~\cite{nickolls2008gpu,jouppi2017tpu}. 
Although GPUs may experience higher per-request overhead due to kernel launches and data movement, they can 
distribute these costs across larger batches, thereby achieving significantly higher throughput for 
inference workloads. Modern CPUs like the \textbf{Granite Rapids} platform bridge this gap by integrating matrix-math 
specialized hardware (AMX) directly into the CPU core, allowing for tensor-heavy execution without the data 
movement penalties of discrete accelerators.

In this study, we explore the practical limits of CPU-based inference by identifying the point at which such 
systems reach performance saturation. Employing a legacy Intel Xeon platform that represents hardware still 
in use in production environments, we establish a clear baseline for inference throughput and latency behavior. 
To provide architectural context, we also reference a modern Xeon server platform, demonstrating how CPU 
architectures have evolved over the past decade. Figure~\ref{fig:xeon_timeline} places the legacy Xeon E5-2403 
v2 used in our experiments alongside a representative modern Granite Rapids-class Xeon within the broader 
architectural progression of Intel Xeon CPUs.

\section{Related Work}
\label{sec:related_work}

This section situates our study within the broader landscape of inference serving, large-scale 
datacenter workloads, and architectural performance modeling.

\subsection{Inference Serving and Tail Latency}
Production inference is generally deployed as an online service where the 
full latency distribution, particularly the high-percentile tail rather than 
simple central tendencies, determines the user experience and adherence to 
service-level objectives. Dean and Barroso 
demonstrate that high-percentile tail latency can dominate end-to-end request time at scale, thereby motivating 
explicit reporting of tail metrics in addition to medians~\cite{dean2013tail}. Clipper 
introduced a practical prediction-serving system that highlights the importance of system-level 
design choices (e.g., batching, caching, and model selection) to meet latency 
targets~\cite{crankshaw2017clipper}. These studies motivate our inclusion of both \textbf{median latency} and 
\textbf{P99 latency} when characterizing CPU-only inference behavior across hardware generations.

\subsection{Benchmarking and Datacenter ML Inference Workloads}
ML inference has emerged as a substantial and growing component of datacenter workloads. 
Hazelwood \emph{et al.} describe the infrastructure and operational constraints of large-scale 
inference deployments and emphasize the role of inference in overall fleet 
utilization~\cite{hazelwood2018facebook}. MLPerf Inference provides a standardized benchmark 
suite and methodology for comparing inference performance across hardware and software 
stacks, and it has become a common reference point for reporting system-level inference 
performance in the literature~\cite{mlperf_inference}. In contrast to broad benchmark 
suites, our study focuses on a controlled, CPU-only baseline on a legacy server platform 
to isolate the scaling effects of batch size and thread-level parallelism.

\subsection{CPU/GPU Performance Modeling and Architectural Limits}
A key question for CPU-only inference is whether performance is fundamentally limited 
by compute throughput or memory-system behavior as workloads scale. The Roofline model 
provides a simple but effective framework to interpret such limits by relating 
operational intensity to peak compute and sustained memory bandwidth~\cite{williams2009roofline}. 
Prior work has also examined how deep learning workloads stress parallel hardware and 
where bottlenecks emerge in practice~\cite{ben2019benchmarking}. 

Complementary efforts explore automated kernel and graph-level optimization to improve 
locality and execution efficiency (e.g., TVM)~\cite{chen2018tvm}, as well as vendor 
toolchains that leverage vectorization and fusion on CPUs (e.g., OpenVINO)~\cite{openvino_intel}. 
Our study builds on these foundations by empirically characterizing the point at which batching and 
threading no longer improve CPU inference throughput on a representative legacy Xeon 
platform, and by using Roofline-style reasoning to interpret the observed saturation behavior.

\section{Deployment Context and Implications}
\label{sec:deployment_context}
\subsection{Operational Resource Constraints}
Beyond architectural limitations, inference deployments in production environments  
often face explicit resource constraints. In many data centers, applications are allocated 
fixed CPU and memory budgets or operate under quota-based resource
management frameworks designed to ensure fair sharing and predictable performance across tenants. These constraints 
are particularly common in legacy or cost-sensitive environments where access to hardware 
accelerators such as GPUs is limited or unavailable. As a result, applications are frequently 
required to use CPU resources conservatively, making aggressive batching or overprovisioning impractical. 
Under such conditions, understanding the performance limits of CPU-only inference becomes crucial 
for capacity planning and latency-sensitive service design.

Quota-based scheduling and resource governance are common in large multi-tenant clusters to 
enforce fair sharing and predictable performance~\cite{verma2015borg}.

\subsection{Widespread Adoption of AI Inference and Capacity Implications}

AI inference has become increasingly prevalent in general-purpose applications, with a growing number of services
incorporating machine learning models for tasks like ranking, 
recommendation, and classification. As a result, pre-trained models are now widely available 
and frequently integrated into application logic, often running on existing CPU-based infrastructure.

This widespread adoption has resulted in a surge of inference workloads competing 
for shared compute resources within data centers. While individual inference tasks may seem 
lightweight from an application perspective, they collectively impose significant aggregate 
demand when deployed at scale. This situation necessitates that data center operators carefully balance 
the mix and capacity of CPU-based servers to accommodate both traditional application logic and 
the rising volume of AI-driven requests.

This balancing act becomes even more challenging due to strict resource governance policies, 
particularly the CPU and memory quotas that are commonly enforced in multi-tenant environments. 
Therefore, understanding the scalability limits of CPU-only inference is crucial for precise capacity 
planning and infrastructure provisioning.

Modern server CPUs are increasingly integrating architectural features designed to enhance machine learning inference 
efficiency, such as wider vector units and specialized instruction support~\cite{intel2024xeon}. 
The rising prominence of AI inference highlights a category of workloads for which traditional CPUs 
were neither initially designed nor thoroughly evaluated.

This shift encourages further research into the behavior of general-purpose processors when subjected to AI-driven 
execution patterns that significantly differ from traditional workloads. Concurrently, modern CPU architectures 
are actively evolving to bridge this gap by increasing vector width, enhancing memory hierarchies, and 
incorporating specialized instruction support for machine learning operations.

As architectural trends evolve, future CPU generations are poised to revolutionize the efficiency with which 
AI workloads are executed on general-purpose hardware. With the rapid and widespread adoption of AI across industries, 
many applications that have long depended on conventional CPU execution are anticipated to undergo significant 
transformations in the coming years. This shift underscores the growing importance of understanding and 
characterizing CPU-based AI inference.
\section{Experimental Setup}

\subsection{Experimental Methodology}
\begin{figure}[t]
    \centering
    \includegraphics[width=\linewidth]{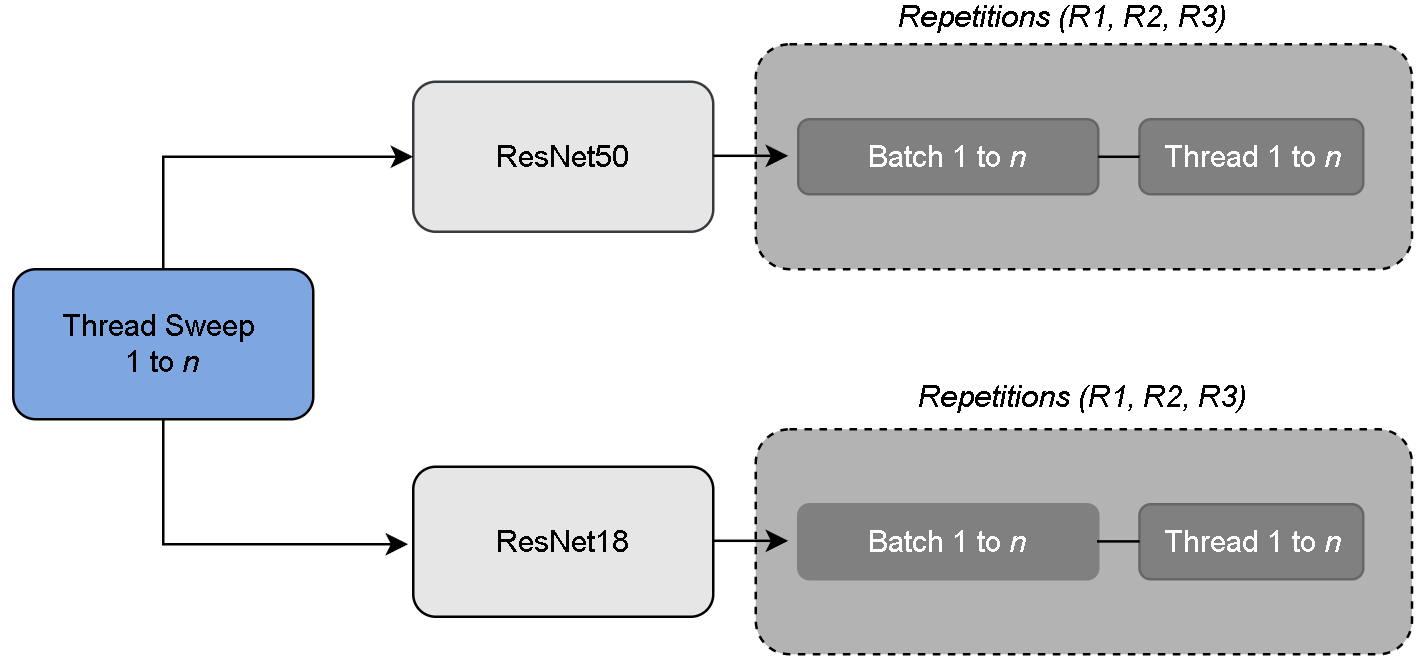}
    \caption{Benchmark execution structure. For each model (ResNet50, ResNet18), we run a sweep over 
	batch sizes and thread counts, repeating each configuration for multiple repetitions (R1--R3) to 
	capture run-to-run variability.}
    \label{fig:thread-sweep-diagram}
\end{figure}

To establish a realistic baseline, we evaluate a legacy server-class CPU, the Intel Xeon E5-2403 v2,
without assistance from hardware accelerators. Although this 4-core processor represents an older
generation, it reflects hardware still commonly deployed in cost-sensitive and long-lived enterprise
environments. To quantify generational scaling and architectural evolution, we additionally evaluate
a modern Intel Xeon Granite Rapids--class processor (Xeon 6 Performance 6521P), representing more than
a decade of CPU microarchitectural advancement.

Our software stack was intentionally kept lean. All experiments were conducted using PyTorch in a
strictly CPU-only configuration to ensure that results reflect raw processor capabilities without
assistance from hardware accelerators.

Thread-level parallelism was controlled using \texttt{torch.set\_num\_threads($t$)} and evaluated independently for each platform.
On the legacy Xeon E5-2403 v2 system, which offers four physical cores without hyper-threading, the number of threads was adjusted
from $t=1$ to $t=4$, aligning precisely with the available hardware parallelism and thereby avoiding oversubscription.
In contrast, on the modern Granite Rapids platform, thread counts were varied from $t=1$ to $t=48$.
CPU affinity was managed via \texttt{taskset};
configurations up to $t=24$ represent scaling across independent physical cores,
whereas configurations from $t=25$ to $t=48$ involve controlled oversubscription onto the same core set.
This oversubscription regime was intentionally included to assess the limits of the processor’s execution pipelines
and memory subsystem under conditions of increasing thread contention.

Table~\ref{tab:hardware_specs} summarizes the key architectural characteristics of the legacy 
Xeon E5-2403 v2 platform and the modern Granite Rapids-class Xeon system evaluated in this 
study.

\subsection{Benchmarking Methodology}

We assess performance by examining both inference latency and throughput. Each experiment follows a structured warm-up 
and measurement procedure to ensure consistent results and to mitigate cold-start effects.

\begin{itemize}
    \item \textbf{Warm-up Phase:} The model performs 20 initial forward passes, which are not recorded. 
    These iterations help initialize the software stack and stabilize hardware caches.
    \item \textbf{Measurement Phase:} We record the individual wall-clock time for 100 consecutive forward passes.
\end{itemize}

The \textbf{latency} ($L$) presented in our results reflects the \textbf{median execution time} per forward pass to ensure robustness against system-level outliers and the performance fluctuations on legacy hardware. For tail analysis, we also report the P99 latency.

The \textbf{throughput} ($P$), measured in images per second (img/sec), indicates the processing capacity for a given batch size $B$ and is calculated using the median latency:
\begin{equation}
    P = \frac{B}{L}
\end{equation}

Given the computational demands of deeper models like ResNet-50, the total execution time for each experimental 
configuration can extend to several minutes on CPU-only systems, ensuring stable and representative average measurements.

We assess CPU inference performance through repeated \emph{thread sweeps}. For each model (ResNet-18 and ResNet-50), 
a sweep evaluates all combinations of batch size and thread count. Figure~\ref{fig:thread-sweep-diagram} illustrates 
the benchmark execution structure, encompassing batch size variation, thread scaling, and repeated measurements.

Thread-level parallelism was managed by explicitly configuring PyTorch intra-op threading through 
\texttt{torch.set\_num\_threads($t$)} prior to each execution. Inter-op parallelism was fixed to a single
thread using \texttt{torch.set\_num\_interop\_threads(1)} to prevent oversubscription. For each
batch--thread configuration, the thread count was held constant to ensure consistent and repeatable
CPU utilization.

\begin{algorithm}[t]
\caption{CPU Inference Benchmark Procedure}
\label{alg:benchmark}
\begin{algorithmic}[1]
\STATE Set \texttt{torch.set\_num\_interop\_threads(1)}
\FOR{platform $\in \{\text{Legacy Xeon}, \text{Granite Rapids Xeon}\}$}
    \IF{platform == Legacy Xeon}
        \STATE threads $\in \{1,2,3,4\}$
    \ELSE
        \STATE threads $\in \{1,2,3,4,6,8,12,16,24,32,40,48\}$
    \ENDIF
    \FOR{model $\in \{\text{ResNet18}, \text{ResNet50}\}$}
        \FOR{threads $\in$ thread set}
            \STATE Set \texttt{torch.set\_num\_threads(threads)}
            \FOR{batch size $\in \{1,2,4,8,16\}$}
                \FOR{repeat $= 1$ to $3$}
					\STATE Run warm-up inference passes
                    \STATE Run inference
                    \STATE Record latency and throughput
                \ENDFOR
            \ENDFOR
        \ENDFOR
    \ENDFOR
\ENDFOR
\end{algorithmic}
\end{algorithm}

Let the set of models be
\begin{equation}
M = \{\text{ResNet18}, \text{ResNet50}\},
\end{equation}
and the set of batch sizes be
\begin{equation}
B = \{1,2,4,8,16\}.
\end{equation}

For the legacy Xeon E5-2403 v2 platform, the thread set is defined as
\begin{equation}
T_{\text{legacy}} = \{1,2,3,4\}.
\end{equation}

For the modern Granite Rapids Xeon 6521P platform, the thread set is defined as
\begin{equation}
T_{\text{granite}} = \{1,2,3,4,6,8,12,16,24,32,40,48\}.
\end{equation}

A single sweep on the legacy platform evaluates all combinations
$(m,b,t) \in M \times B \times T_{\text{legacy}}$, yielding
\begin{equation}
|M| \times |B| \times |T_{\text{legacy}}|
= 2 \times 5 \times 4 = 40
\end{equation}
unique configurations.

A single sweep on the Granite Rapids platform evaluates all combinations
$(m,b,t) \in M \times B \times T_{\text{granite}}$, yielding
\begin{equation}
|M| \times |B| \times |T_{\text{granite}}|
= 2 \times 5 \times 12 = 120
\end{equation}
unique configurations.

Each configuration is executed $R=3$ times to account for run-to-run variability,
resulting in a total of 120 executions on the legacy platform and 360 executions
on the Granite Rapids platform.

Individual inference executions occur per thread sweep.
The three repetitions are aggregated to yield a single aggregated measurement
for each configuration. For each configuration, we report the median latency and
throughput across the three repetitions, with variability quantified using the
sample standard deviation.

In addition to the $R=3$ repetitions per configuration within a sweep, we repeat
the entire thread sweep $S=10$ times to capture run-to-run variability across
independent benchmark passes. Thus, each configuration is executed a total of
$R \times S = 30$ times. Unless otherwise noted, reported means and standard
deviations are computed across all $R \times S$ executions.

\begin{table}[t]
\centering
\caption{Hardware Specifications of Legacy and Modern Server CPUs}
\label{tab:hardware_specs}
\begin{tabular}{lcc}
\hline
\textbf{Specification} & \textbf{Xeon E5-2403 v2 (legacy)} & \textbf{Xeon 6 Perf (modern)} \\
\hline
Microarchitecture & Ivy Bridge-EN & Granite Rapids \\
Process Tech      & 22\,nm        & Intel 3 \\
Release Year      & 2013          & 2024--2025 \\
Cores / Threads   & 4 / 4         & 24 / 48 \\
Base Frequency    & 1.80\,GHz     & 2.60\,GHz \\
Max Turbo         & N/A           & Enabled \\
LL Cache          & 10\,MB (L3)   & 144\,MB (LLC) \\
Memory Type       & DDR3 (8\,GB)  & DDR5-6400 (32\,GB) \\
Memory Speed      & 1600\,MT/s    & 6400\,MT/s \\
TDP               & 80\,W         & 225\,W \\
\hline
\end{tabular}
\end{table}

\subsection{Hardware Platform}

The experiments were conducted on two server-class CPU platforms, representing both legacy and 
modern enterprise deployments. The legacy system is equipped with an Intel Xeon E5-2403 v2 processor, 
operating at 1.80~GHz, and features four physical cores without simultaneous multithreading. 
This server is provisioned with 8~GB of system memory and runs on a Linux-based operating system.

In contrast, the modern platform is equipped with an Intel Xeon Granite Rapids--class processor
(Xeon 6 Performance 6521P), featuring 24 physical cores with simultaneous multithreading enabled
(48 threads) and 32\,GB of DDR5-6400 system memory. This platform represents a contemporary
high-core-count server architecture designed for throughput-oriented and AI-driven workloads.
Detailed architectural characteristics for both systems are summarized in
Table~\ref{tab:hardware_specs}.

\subsection{Software Environment}

Inference was conducted using the PyTorch~\cite{paszke2019pytorch} deep learning framework in CPU-only
mode, with CUDA explicitly disabled. The software runtime configuration is summarized in Table~\ref{tab:sw_runtime}. 
The legacy server executed workloads directly on a CentOS~7.5-based
environment, reflecting long-term enterprise deployments. Meanwhile, the modern Granite Rapids platform ran the
experiments within an Ubuntu~20.04 Linux container (LXC) hosted on a newer Linux distribution. This setup 
preserved compatibility with the legacy Python and library stack while maintaining identical runtime
dependencies across platforms. This approach minimizes confounding effects associated with differences in operating
system and library, facilitating fair architectural comparison and reproducibility across
heterogeneous hardware.

\begin{table}[t]
\caption{Software and Runtime Configuration}
\label{tab:sw_runtime}
\centering
\begin{tabular}{p{3.5cm} p{4.5cm}}
\hline
\textbf{Category} & \textbf{Setting} \\
\hline
OS / Kernel &
\textbf{Legacy:} CentOS Linux 7.5 (kernel 3.10); 
\textbf{GNR:} Ubuntu 20.04 (LXC) on Linux host.\\

Python / PyTorch &
Python 3.6.8, PyTorch 1.10.1 (CPU-only). \\

Compiler / Libraries &
GCC v4.8.5 / 9.4.0; OpenMP (libgomp) with Intel MKL and oneDNN enabled. \\

Thread Control & Intra-op: \texttt{set\_num\_threads(t)}; \newline Inter-op: fixed to 1 via \texttt{set\_num\_interop\_threads(1)}. \\

CPU Affinity & \textbf{Legacy:} Unpinned; \textbf{GNR:} Pinned to physical cores (0--23), excluding SMT siblings. Thread counts exceeding the core count therefore reflect controlled oversubscription onto the same core set. \\

\hline
\end{tabular}
\end{table}

Table~\ref{tab:sw_runtime} summarizes the software stack, runtime configuration, and threading controls used 
in all experiments.

\subsection{Models and Inference Configuration}
We evaluated inference performance using standard convolutional neural network architectures from the ResNet 
family. All models were run in inference mode with gradient computation disabled. Batch size was systematically varied 
while all other parameters were held constant. Unless otherwise specified, the results presented isolate batching effects by 
setting the thread count to one; multi-thread scaling is analyzed separately via thread sweeps.

All models evaluated in this study were pre-trained on the ImageNet dataset~\cite{deng2009imagenet}.

\begin{figure*}[t]
  \centering
  \begin{subfigure}[t]{0.48\textwidth}
    \centering
    \includegraphics[width=\linewidth]{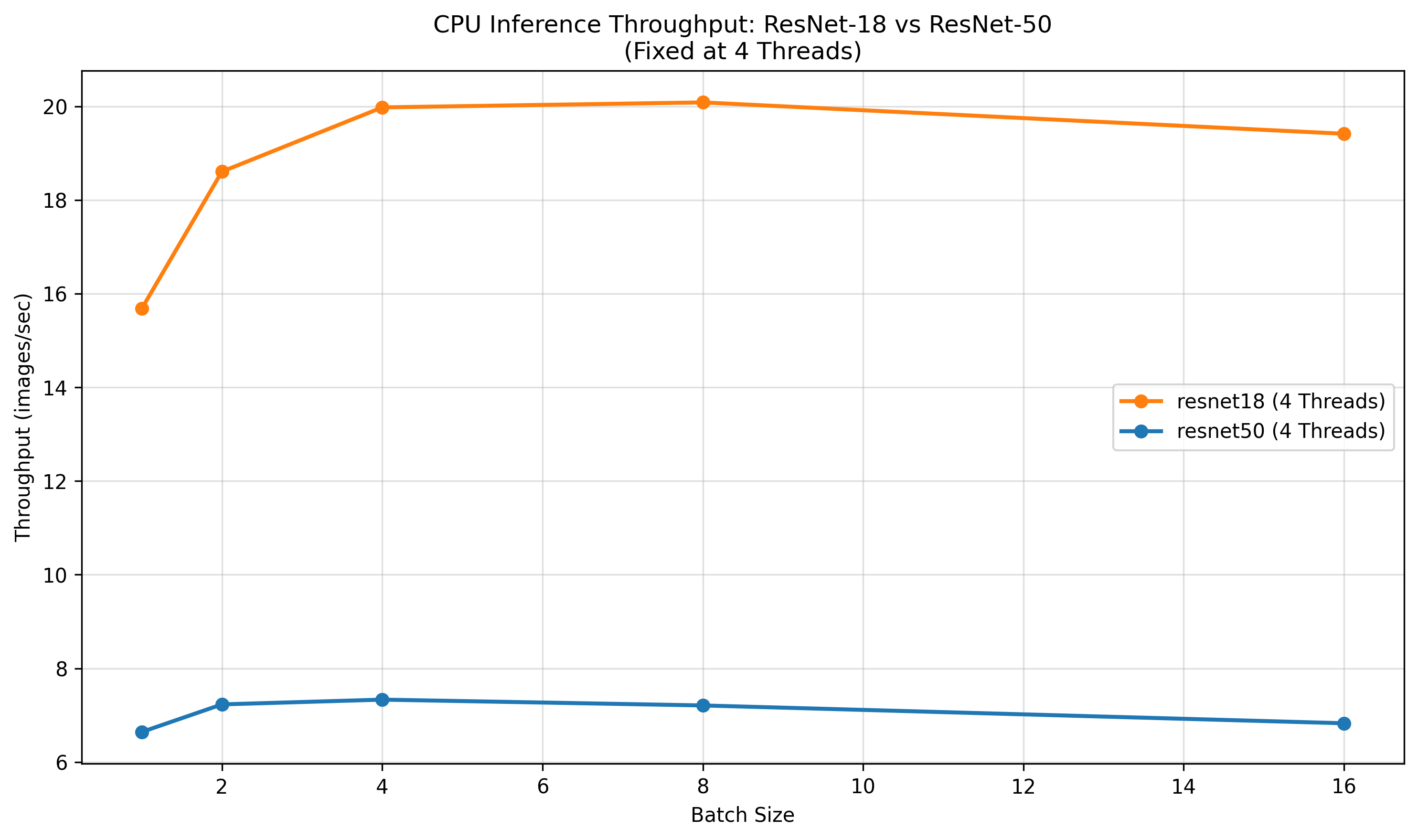}
    \caption{Legacy Server (4 Threads): Median throughput saturates by $B=4$, with changes of less than $0.2\%$ between $B=4$ and $B=8$, indicating a flat response within measurement variability. ResNet-18 achieves $\sim3\times$ higher throughput than ResNet-50 due to architectural limits of the legacy platform.}
  \label{fig:throughput_legacy}
  \end{subfigure}
  \hfill
  \begin{subfigure}[t]{0.48\textwidth}
    \centering
    \includegraphics[width=\linewidth]{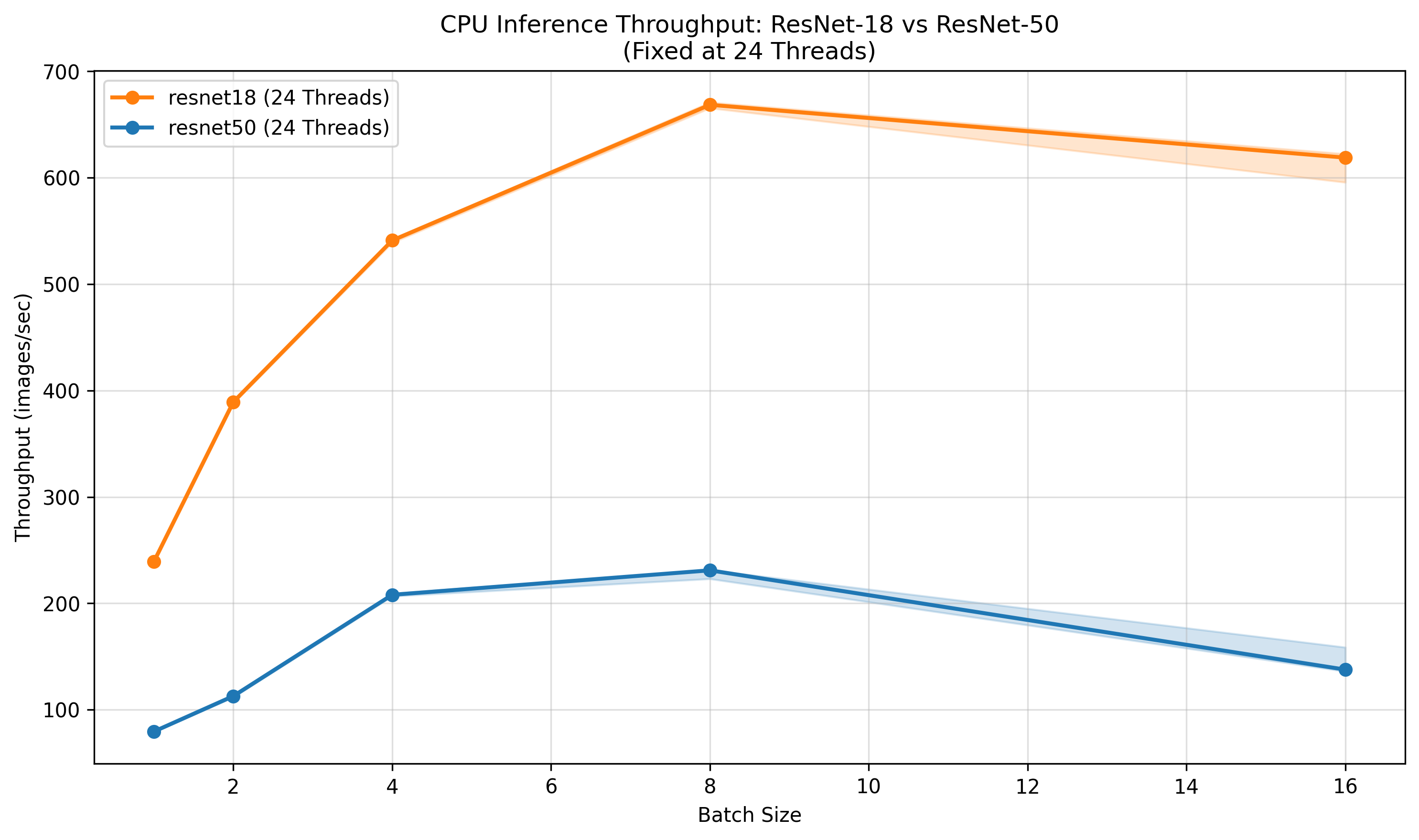}
    \caption{Modern Server (24 Threads): Median throughput peaks at batch size 8, reaching $>600$ images/sec, representing a $\sim30\times$ generational improvement.}
    \label{fig:throughput_modern}
  \end{subfigure}

  \caption{Peak-to-peak CPU Inference Throughput Comparison: Evaluation of ResNet architectures across legacy (4-thread) and modern (24-thread) environments. To ensure execution determinism and filter system jitter, all trend lines report the median images per second (IPS). The modern server is pinned to 24 physical cores to avoid hyperthreading-induced performance cliffs, representing a comparison of optimized system capacities across hardware generations.}
  \label{fig:full_throughput_comparison}
\end{figure*}

\section{CPU Inference Throughput Scaling}

Figure~\ref{fig:full_throughput_comparison} illustrates a peak-to-peak comparison of CPU inference throughput as batch size increases from 1 to 16. To evaluate each platform at its optimized capacity, the legacy platform is tested at its hardware limit of 4 physical cores, while the Granite Rapids platform is evaluated at a 24-thread load pinned to physical cores.

On the legacy Xeon platform, throughput improves modestly as the batch size increases from 1 to 4 but saturates quickly by batch size 4, remaining flat through batch size 8. For ResNet-50, performance plateaus at a median of $7.33$~IPS, while ResNet-18 reaches a median of $20.08$~IPS at a batch size of 8. At larger batch sizes (12 and 16), throughput slightly declines, indicating early saturation of the limited 4-core resources and DDR3 memory bandwidth.

In contrast, the Granite Rapids platform demonstrates significantly higher scalability. Throughput continues to scale effectively up to a batch size of 8. At this peak configuration, the modern server achieves a median throughput of $230.98$~IPS for ResNet-50 and $668.58$~IPS for ResNet-18---representing a \textbf{31.5$\times$} and \textbf{33.3$\times$} peak-to-peak performance increase over the legacy system, respectively. On both platforms, throughput begins to regress at a batch size of 16, suggesting that memory bandwidth limits or management overhead begin to outweigh the benefits of increased parallelism.

\section{Latency Behavior and Variability}

\subsection{Architectural Performance Under Hardware Parity and Full-Core Utilization}

\begin{figure*}[ht]
  \centering
  \begin{subfigure}[t]{0.48\textwidth}
    \centering
    \includegraphics[width=\textwidth]{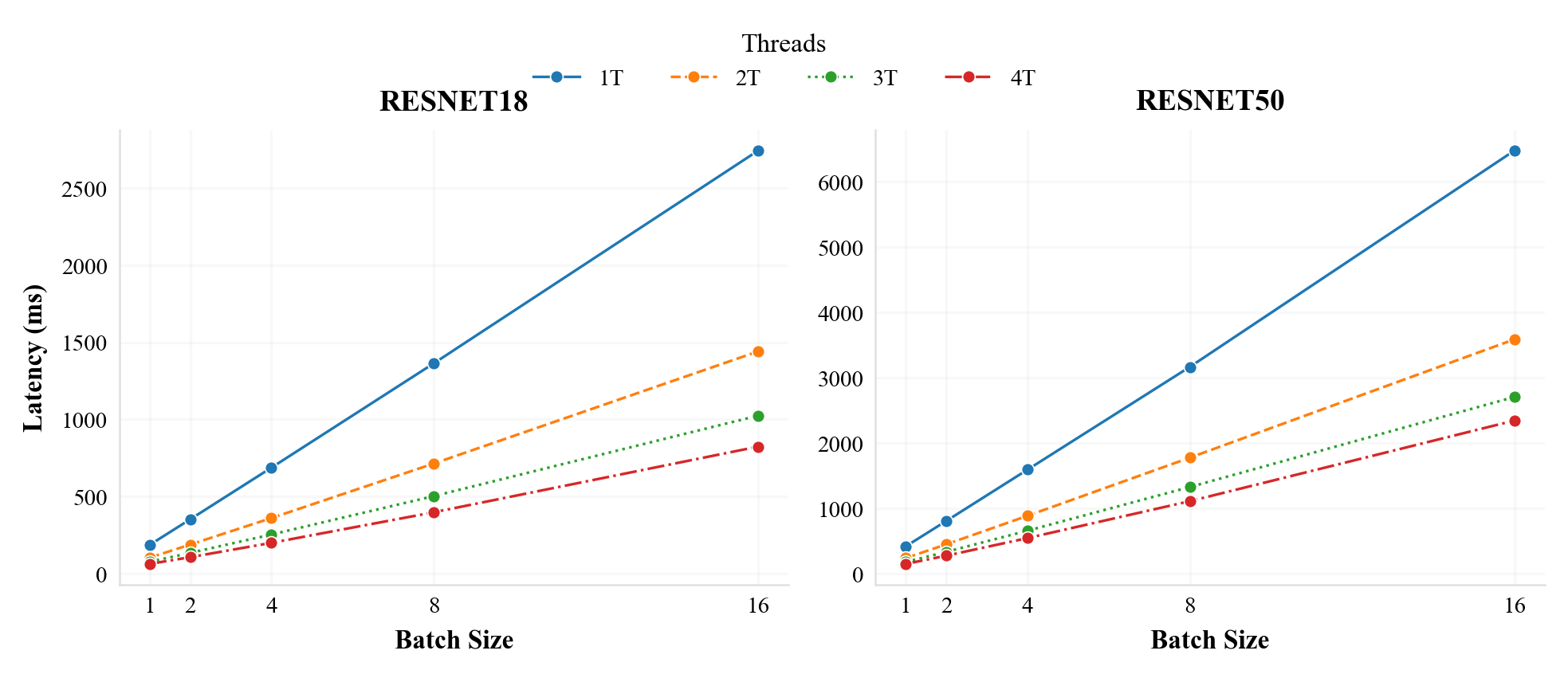}
    \caption{Legacy Xeon server: CPU inference latency as a function of batch size under constrained parallelism.
Results are shown for thread counts ranging from $1$ to $4$.
Single-threaded execution (1T) exhibits a sharp increase in latency as batch size grows, with ResNet-50
reaching approximately $6{,}500$~ms at the largest evaluated batch size, highlighting the limitations of
legacy CPU architectures under batched inference workloads.}
\label{fig:latency_legacy_sweep}
  \end{subfigure}
\hfill
  \begin{subfigure}[t]{0.48\textwidth}
    \centering
    \includegraphics[width=\textwidth]{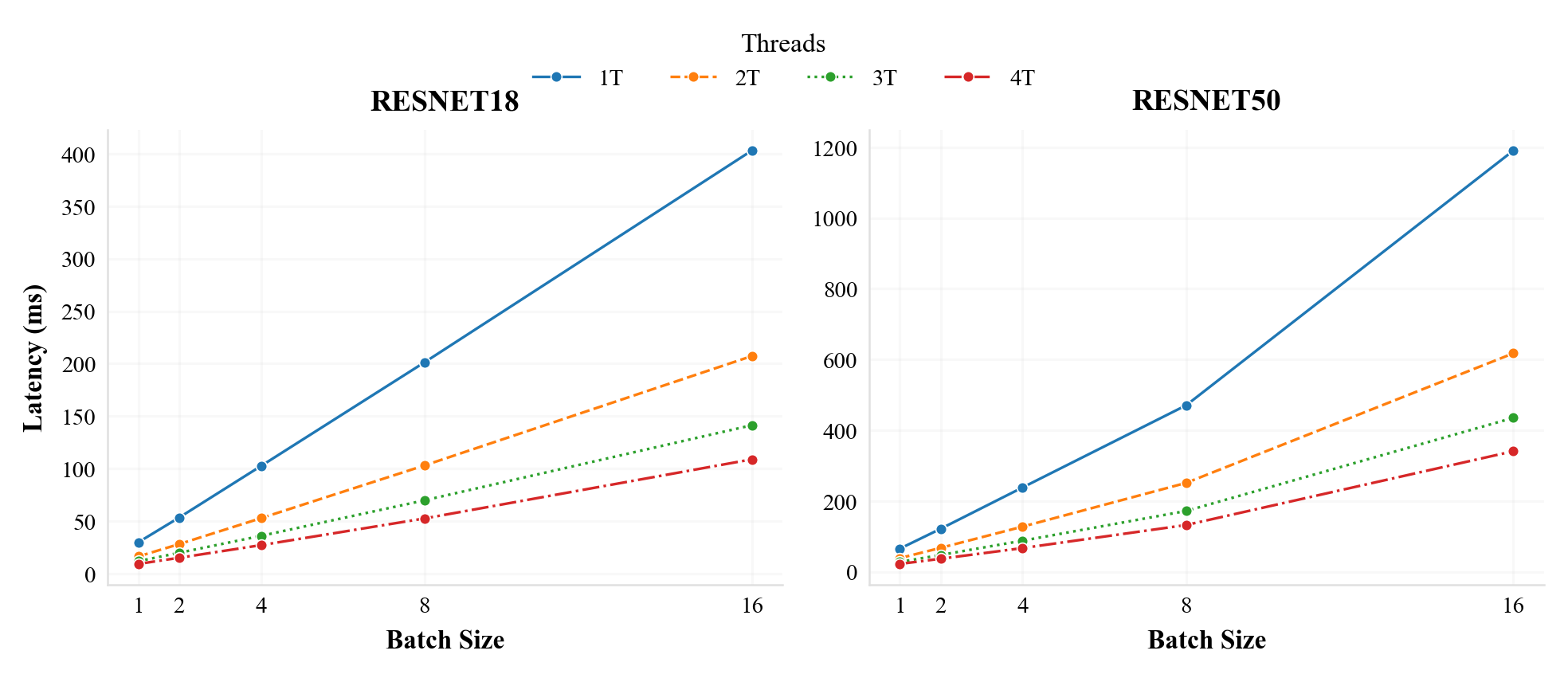}
    \caption{Modern Granite Rapids server: CPU inference latency as a function of batch size under constrained
parallelism.
Results are shown for thread counts ranging from $1$ to $4$.
Latency increases smoothly with batch size across all configurations, with single-threaded execution (1T)
remaining below approximately $1{,}200$~ms at the largest evaluated batch size.}
\label{fig:latency_modern_sweep}
  \end{subfigure}

\vspace{12pt}

\begin{subfigure}[t]{0.48\textwidth}
    \centering
    \includegraphics[width=\textwidth]{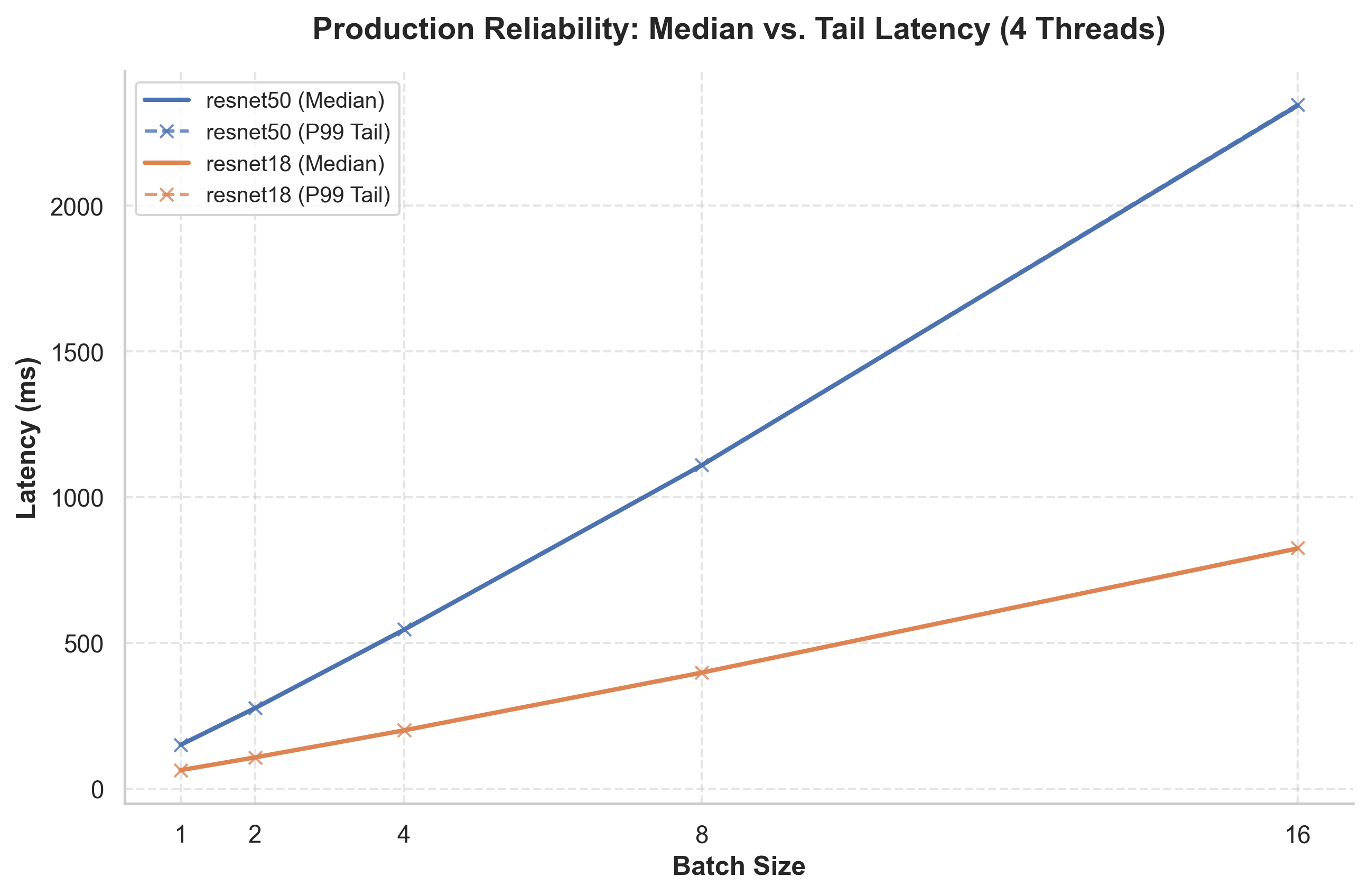}
    \caption{Legacy server tail analysis: Median and P99 latency at a standardized $4$-thread configuration.
    P99 latency closely tracks the median across batch sizes, indicating limited tail amplification.}
    \label{fig:tail_latency_legacy}
\end{subfigure}
\hfill
\begin{subfigure}[t]{0.48\textwidth}
    \centering
    \includegraphics[width=\textwidth]{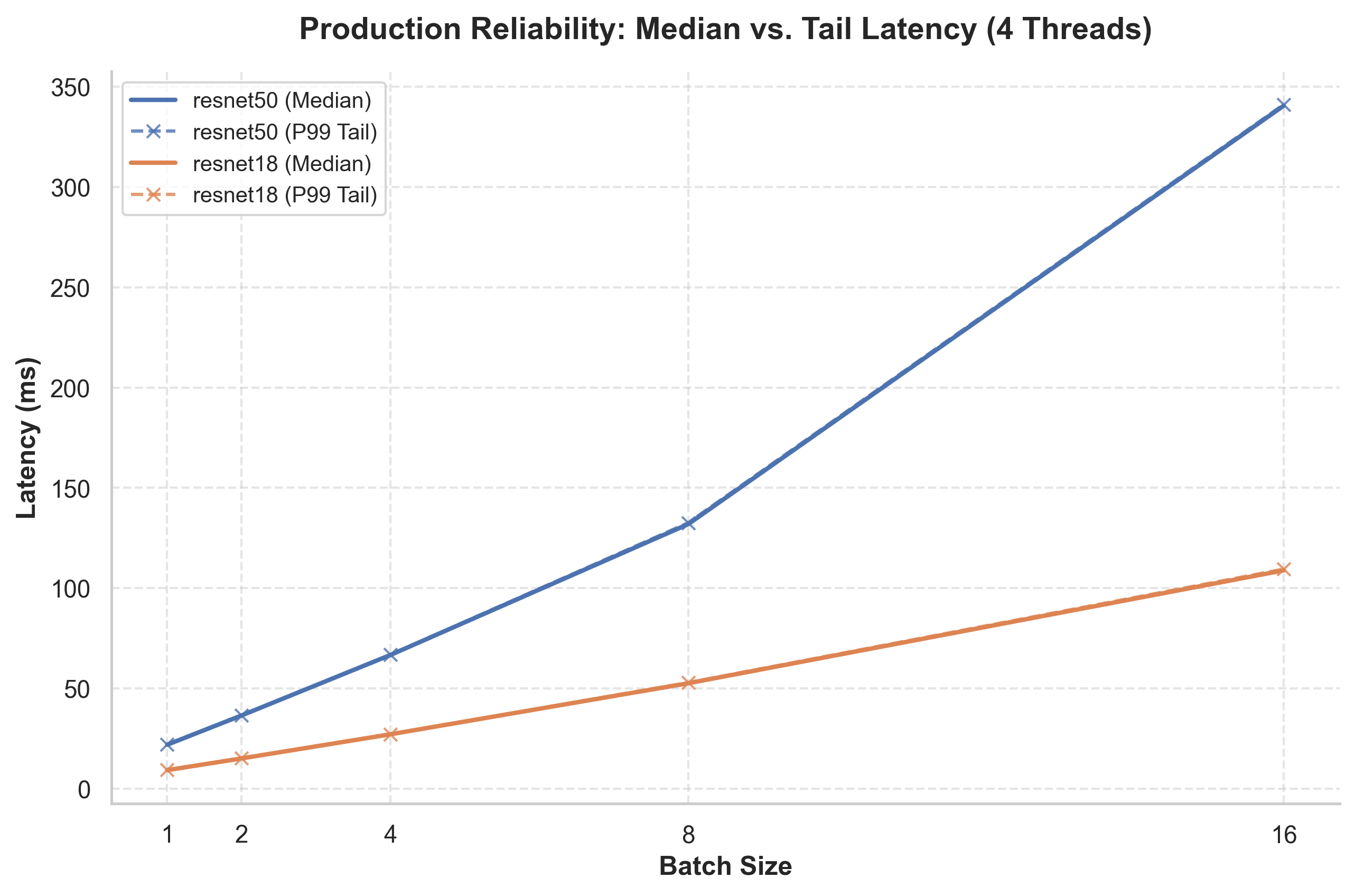}
    \caption{Modern server tail analysis: Median and P99 latency at a standardized $4$-thread configuration.
    Median and tail latency remain closely aligned across batch sizes, reflecting stable execution.}
    \label{fig:tail_latency_modern}
\end{subfigure}

 \caption{CPU inference latency characteristics comparing the legacy Xeon and modern Granite Rapids platforms.
\textbf{Top row:} latency as a function of batch size under constrained parallelism, with thread counts limited
to $1$--$4$ on both platforms to maintain hardware parity.
\textbf{Bottom row:} tail latency analysis (median vs.\ P99) conducted at a fixed $4$-thread configuration,
highlighting production reliability under identical execution resources.}
\label{fig:full_latency_comparison}
\end{figure*}

The generational performance gap is most evident in the scaling of inference
latency with batch size. As shown in
Figure~\ref{fig:latency_legacy_sweep}, the legacy Xeon architecture exhibits
severe latency amplification \textbf{under single-threaded ($1T$) execution}.
In this unoptimized configuration, ResNet-50 latency increases nearly
linearly with batch size, reaching approximately 6{,}500~ms at $B=16$.

However, this cliff is substantially mitigated through parallelism. To provide a rigorous architectural comparison based on \textbf{hardware parity}, Figure~\ref{fig:latency_modern_sweep} evaluates the modern Granite Rapids architecture by constraining its execution to a standardized 4-thread baseline. By limiting the modern server to the same thread count as the legacy system's maximum capacity, we isolate the performance gains attributable to core efficiency and cache hierarchy rather than raw core count. 

\begin{figure*}[t]
    \centering

    \begin{subfigure}[b]{0.32\textwidth}
        \centering
        \includegraphics[width=\textwidth]{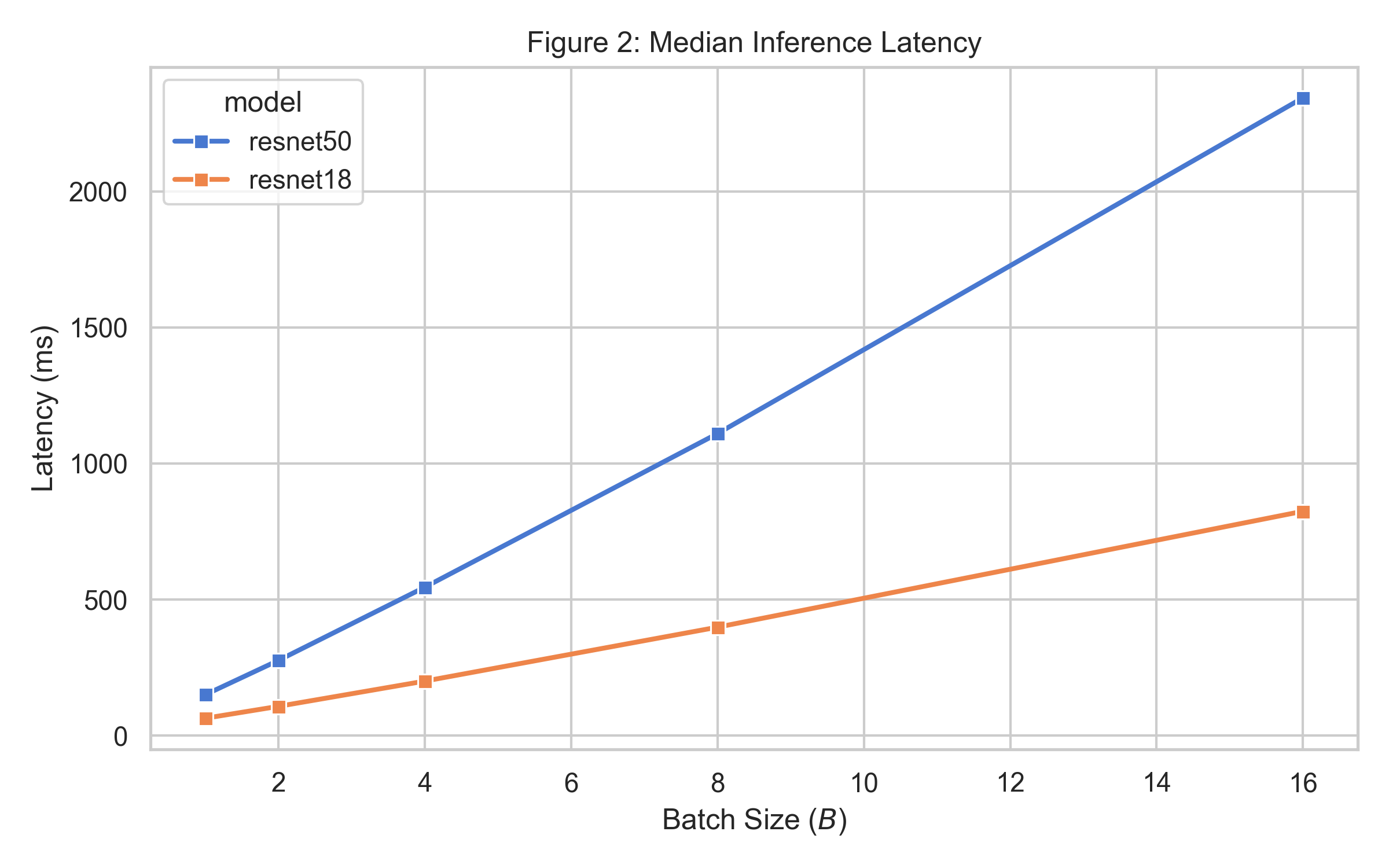}
        \caption{Median Latency}
        \label{fig:latency_legacy}
    \end{subfigure}
    \hfill
    \begin{subfigure}[b]{0.32\textwidth}
        \centering
        \includegraphics[width=\textwidth]{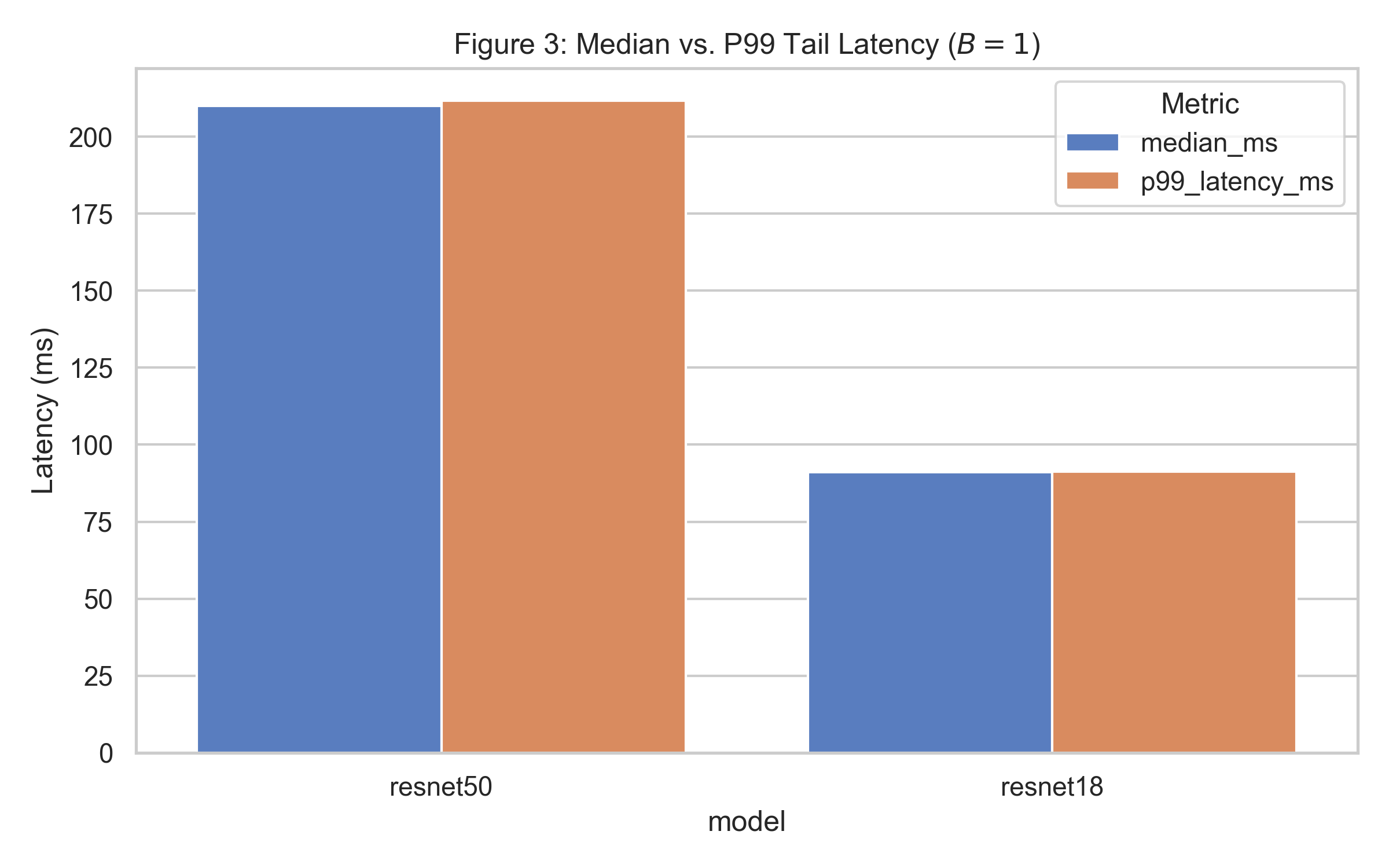}
        \caption{Tail Latency (P99)}
        \label{fig:tail_legacy}
    \end{subfigure}
    \hfill
    \begin{subfigure}[b]{0.32\textwidth}
        \centering
        \includegraphics[width=\textwidth]{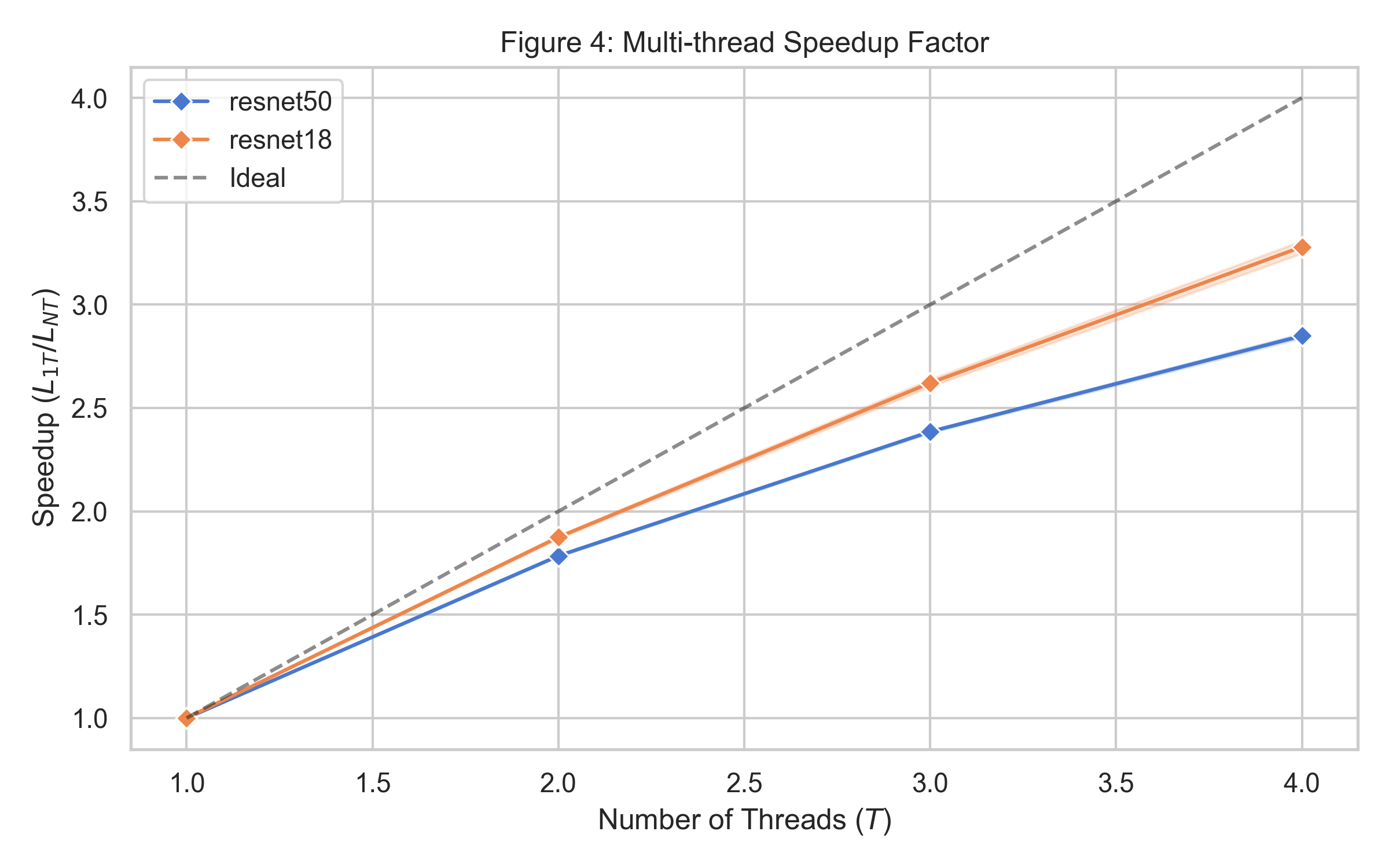}
        \caption{Parallel Speedup}
        \label{fig:speedup_legacy}
    \end{subfigure}

    \vspace{12pt} 


    \begin{subfigure}[b]{0.32\textwidth}
        \centering
        \includegraphics[width=\textwidth]{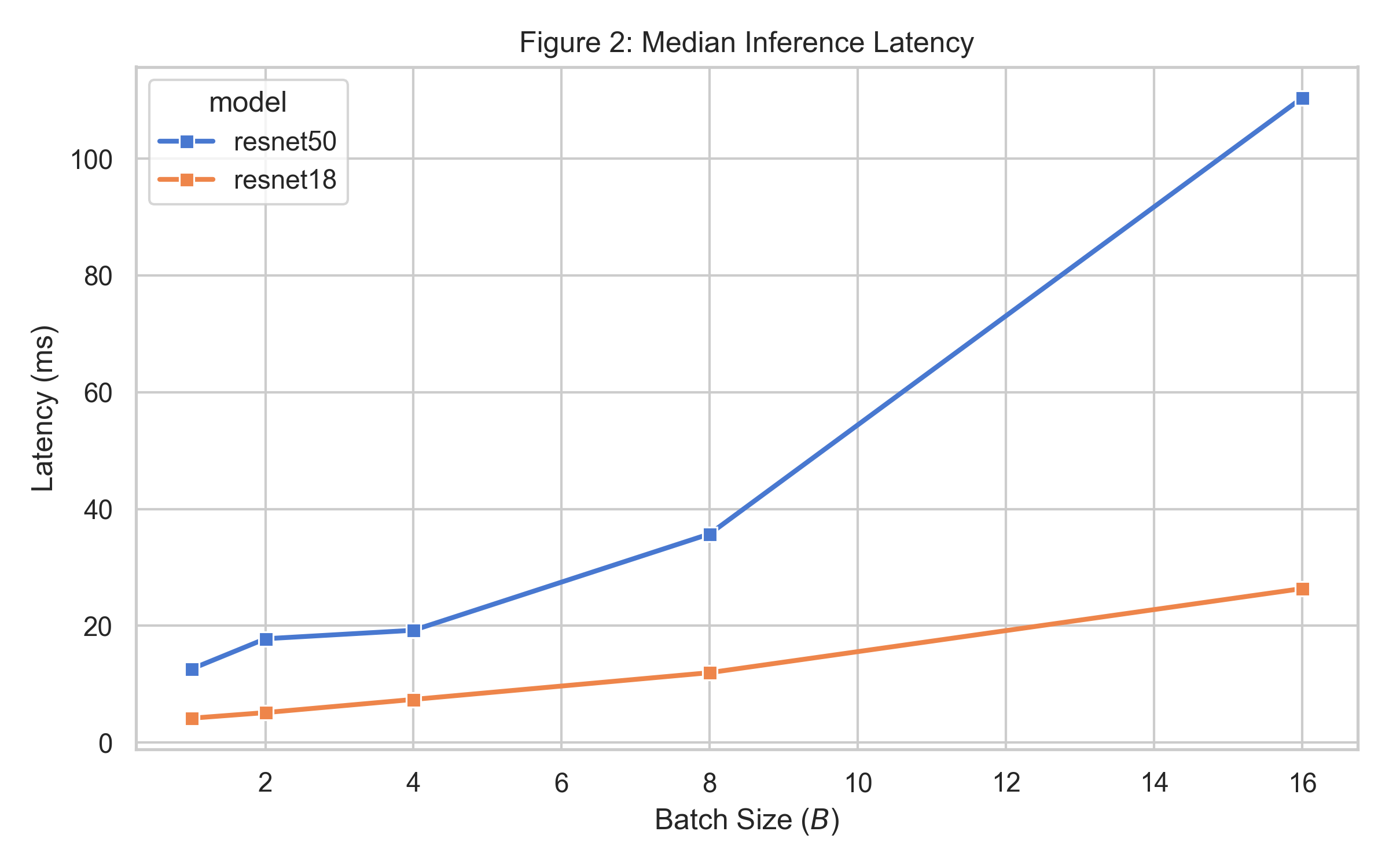}
        \caption{Median Latency}
        \label{fig:latency_modern}
    \end{subfigure}
    \hfill
    \begin{subfigure}[b]{0.32\textwidth}
        \centering
        \includegraphics[width=\textwidth]{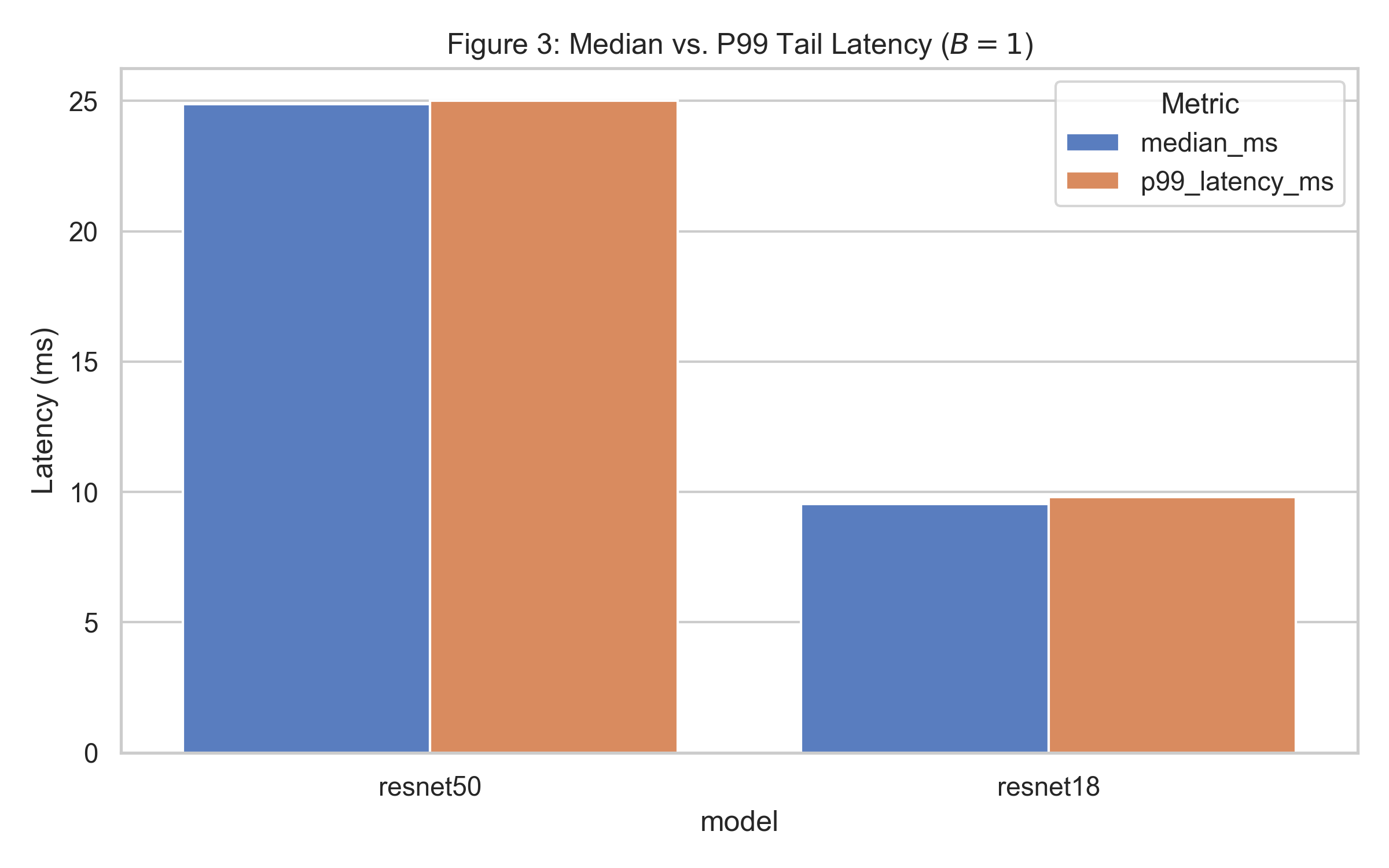}
        \caption{Tail Latency (P99)}
        \label{fig:tail_modern}
    \end{subfigure}
    \hfill
    \begin{subfigure}[b]{0.32\textwidth}
        \centering
        \includegraphics[width=\textwidth]{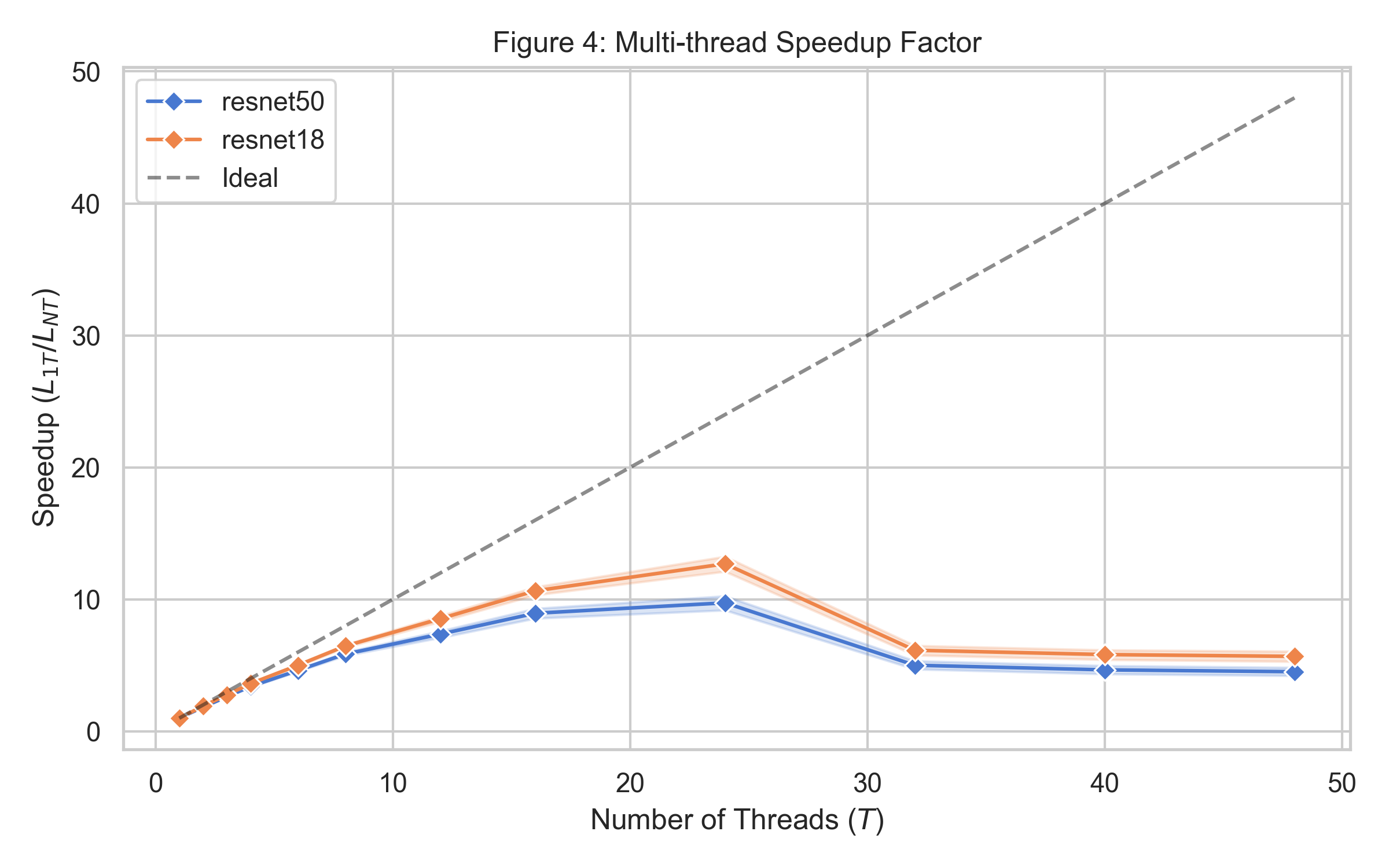}
        \caption{Parallel Speedup}
        \label{fig:speedup_modern}
    \end{subfigure}

\caption{Comparative inference performance of ResNet-18 and ResNet-50 across CPU generations.
The top row shows results for the legacy Intel Xeon E5-2403 v2 evaluated at its maximum capacity of 4 physical cores,
while the bottom row presents results for the modern Intel Xeon Granite Rapids evaluated at its maximum capacity of 24 physical cores.
Columns, from left to right, report (a,d) median inference latency as a function of batch size,
(b,e) median versus P99 tail latency at $B=1$, and (c,f) multi-thread speedup relative to single-thread execution.
Latency values are reported as medians across independent sweeps to mitigate the influence of outliers.
Speedup results extend beyond the physical core count to include logical cores, exposing the effects of oversubscription and shared-resource contention.
Overall, the figure highlights the contrasting scalability limits and architectural behavior of legacy and modern CPU platforms under full utilization.}

\label{fig:full_performance_comparison}
\end{figure*}

Even under these parity-constrained conditions, the modern architecture demonstrates superior scaling:
\begin{itemize}
    \item \textbf{Mitigation of Scaling Penalties:} While the legacy system requires multi-threading ($4T$) to bring $B=16$ ResNet-50 latency down to $\approx$2,300~ms, the modern architecture achieves a significantly lower $\approx$1,200~ms using only a single thread ($1T$).
    \item \textbf{Parallel Scaling at Parity:} At the 4-thread parity baseline, the modern server achieves a $B=16$ median latency of $\approx$350~ms, representing a $6.5\times$ improvement over the legacy peak of $\approx$2,300~ms.
    \item \textbf{Median Latency Divergence (Full-Core Utilization):} 
Under fully optimized configurations utilizing all available physical cores, the modern platform maintains a median ResNet-50 latency of only 116~ms (Figure~\ref{fig:latency_modern}), compared to the legacy platform’s median peak of approximately 2,300~ms (Figure~\ref{fig:latency_legacy}). This stark divergence underscores the generational leap in architectural capability for CPU-based AI inference.
\end{itemize}

The following series of plots (Figures~\ref{fig:latency_legacy}--\ref{fig:speedup_modern}) 
summarizes the throughput scaling, median latency trends, tail-latency 
stability, and thread-scaling efficiency across both platforms.

The findings indicate that legacy hardware is often inadequate for real-time CNN inference workloads at moderate batch sizes, whereas the modern architecture provides a feasible solution for CPU-based deep learning inference.

Figure~\ref{fig:latency_legacy} shows inference latency as a function of batch size for both the legacy and modern platforms under full physical-core utilization, reflecting the maximum practical capacity of each system in deployment.

Latency increases with batch size for both ResNet-18 and ResNet-50 across both architectures. While batching amortizes per-inference overhead, larger batch sizes require the CPU to process substantially larger activation tensors and intermediate feature maps, resulting in increased memory traffic and cache pressure.

However, the absolute latency values differ by more than an order of magnitude.
On the legacy Xeon E5-2403 v2, throughput saturates at low batch sizes
($B=4$--$8$), after which median latency continues to increase sharply with
additional batching, reaching $\approx$2300~ms at $B=16$.
In contrast, on the Granite Rapids (GNR) platform, throughput saturates at
$B=8$ using 24 physical cores, with a median per-inference latency of
$\approx$35~ms at this operating point. Even at larger batch sizes
($B=16$), latency remains substantially lower at $\approx$110~ms,
representing a $\sim$21$\times$ reduction relative to the legacy platform.

As the batch size $B$ increases, the memory footprint of intermediate activation tensors grows, increasing working-set size. Given the legacy platform's 10\,MB shared L3 cache and DDR3 memory, the observed throughput plateau beyond $B>4$ is consistent with rising cache pressure and main-memory bottlenecks. In contrast, the Granite Rapids server leverages a massive 144\,MB Last-Level Cache (LLC) and high-bandwidth DDR5-6400 memory. This allows the modern platform to contain much larger working sets within high-speed cache, delaying the onset of memory-bound stalls and enabling effective throughput scaling up to $B=8$.

Importantly, the increase in latency on the legacy platform is not matched by proportional throughput improvements. Gains quickly diminish as the processor becomes limited by narrow memory bandwidth. 
This behavior highlights a fundamental trade-off: on legacy platforms, batching improves utilization at the cost of impractical response times. 
On the Granite Rapids platform, however, the trade-off is significantly shifted; 
the high baseline performance allows for larger batch sizes (e.g., $B=8$) to 
provide maximum throughput while still maintaining median latencies 
(\textbf{35~ms} for ResNet-50, as shown in Figure~\ref{fig:latency_modern}) that remain viable for near-real-time inference.

Across all configurations and both platforms, latency variance remains remarkably low. This consistency suggests that latency inflation with increased batch size is an inherent property of CPU-based inference. However, while the Legacy server is constrained by its 4-thread limit, the Modern server—featuring 24 physical cores and 48 logical threads—demonstrates how architectural advancements in memory subsystems significantly mitigate the magnitude of this penalty, even when restricted to the same 4-thread execution envelope.

\subsection{Throughput and Batch Efficiency}
As illustrated in Figure~\ref{fig:throughput_legacy}, ResNet-18 consistently
outperforms ResNet-50 on the legacy Xeon platform. At saturation
($B=4$--$8$), ResNet-18 achieves a median throughput of approximately
20~Images Per Second (IPS), compared to approximately 7~IPS for ResNet-50,
corresponding to a $\sim$2.7$\times$ throughput advantage.

On the Granite Rapids (GNR) platform, this performance gap widens slightly.
At batch size $B=1$, ResNet-18 achieves a median throughput of approximately
230~images/s, compared to approximately 80~images/s for ResNet-50,
corresponding to a $\sim$3$\times$ throughput advantage.

A critical observation is the divergence in batch efficiency between the two architectures.
The legacy Xeon platform exhibits a pronounced flat-line behavior: for ResNet-50, throughput
improves only marginally from $B=1$ to its saturation region ($B=4$--$8$), and remains effectively
unchanged through $B=16$. This indicates that the legacy system reaches an architectural
throughput ceiling at very low batch sizes, where additional batching cannot overcome the
constraints imposed by its 4-core design and limited DDR3 memory bandwidth.

\begin{table}[ht]
\centering
\caption{Peak-to-Peak Median Throughput Comparison (IPS)}
\label{tab:median_throughput}
\begin{tabular}{lccc} 
\toprule
\textbf{Model} & \textbf{\begin{tabular}[c]{@{}c@{}}Legacy Xeon\\ (BS 4 Saturates)\end{tabular}} & \textbf{\begin{tabular}[c]{@{}c@{}}Granite Rapids\\ (BS 8 Peak)\end{tabular}} & \textbf{Speedup} \\
\midrule
ResNet-18 & 20.08 & 668.58 & 33.3$\times$ \\
ResNet-50 & 7.33  & 230.98 & 31.5$\times$ \\
\bottomrule
\end{tabular}
\end{table}

In contrast, the Granite Rapids platform demonstrates substantial scaling gains.
For ResNet-50, throughput increases from 79.6~IPS at $B=1$ to a peak of
230.9~IPS at $B=8$, representing a \textbf{2.9$\times$ improvement} achieved
through batching alone under a fixed 24-thread configuration. This scaling
behavior is enabled by the modern architecture’s large 144~MB last-level cache
and DDR5 memory subsystem, which together allow computation and memory traffic
to be effectively amortized across all 24 physical cores.

Ultimately, the performance gap between the two generations is pronounced:
at their respective peak throughput configurations, the Granite Rapids server
achieves a \textbf{31.5$\times$ speedup} for ResNet-50 compared to the legacy
Xeon platform (Table~\ref{tab:median_throughput}). Furthermore, the high
reproducibility of these measurements across repeated runs on both platforms
indicates that the observed throughput characteristics are stable and
repeatable. This confirms that the stagnation of the legacy platform and the
scaling behavior of the modern architecture reflect fundamental hardware
constraints rather than transient software effects.

\subsection{Latency Scaling and Computational Cost}

Figures~\ref{fig:latency_legacy} and~\ref{fig:latency_modern} show an approximately linear
relationship between batch size ($B$) and \textbf{median inference latency} on both platforms.
On the legacy Xeon, ResNet-50 median latency increases from 210~ms at $B=1$ to approximately
2{,}300~ms at $B=16$. In contrast, the Granite Rapids (GNR) platform completes the same
single-threaded $B=1$ inference in 23~ms, reaching approximately 210~ms at $B=16$.

Multi-threaded execution substantially reduces absolute latency on both systems.
When utilizing all available physical cores (24 on GNR), the GNR platform achieves a
$B=16$ \textbf{median latency} of approximately 210~ms, whereas the legacy platform
remains dominated by significantly higher execution overhead.
In latency-sensitive production environments, where service-level objectives (SLOs)
often require sub-200~ms response times~\cite{dean2013tail}, the legacy platform is
effectively constrained to $B=1$, while the Granite Rapids system can support larger
batch sizes within interactive limits.

This behavior underscores that batching within the legacy architecture results in 
increased latency without corresponding improvements in throughput.

\subsection{Tail Latency and System Jitter}
Figure ~\ref{fig:tail_legacy} and ~\ref{fig:tail_modern} compares the median latency with the $99^{\text{th}}$ percentile (P99) 
tail latency. The P99 metric is a critical indicator of system reliability in production 
environments. Our measurements indicate that for both platforms, the P99 latency closely 
tracks the median across all tested batch sizes. 

At $B=1$, the variance is nearly negligible on the legacy Xeon platform:
ResNet-50 exhibits a median latency of approximately 210~ms with a closely
tracking P99 tail of approximately 212~ms. For the lighter ResNet-18 model,
median and P99 latencies remain tightly coupled at approximately 90~ms.
On the Granite Rapids (GNR) platform, 
this deterministic behavior is equally pronounced; ResNet-50 exhibits a median latency 
of approximately 24.8~ms with a P99 of approximately 25.0~ms, while ResNet-18 shows 
a median latency of approximately 9.5~ms with a closely tracking P99 tail.

As the workload increases to $B=8$, the alignment remains tight. On the Granite
Rapids (GNR) platform, the P99/median margin remains under 6\%, with a median
latency of 174.3~ms compared to a P99 of 184.7~ms. Even on the legacy Xeon,
where throughput saturates at low batch sizes, tail latency does not decouple
significantly from the median. However, it is important to note that while the
\textit{relative} jitter remains low, the \textit{absolute} latency on legacy
hardware under single-threaded ($1T$) execution continues to grow sharply,
reaching approximately 6{,}500~ms at $B=16$.

\subsection{Multi-threaded Scalability Analysis}
Figure~\ref{fig:speedup_legacy} and \ref{fig:speedup_modern} illustrates the speedup factor, defined as 
$S = L_{1T} / L_{NT}$. On the legacy Xeon E5-2403 v2 platform (comprising 4 physical cores), 
the speedup curve deviates from the ideal linear scaling almost immediately, 
reaching a modest maximum of $3.28\times$ at the 4-thread limit. 

In contrast, the Granite Rapids (GNR) platform demonstrates significantly greater scaling 
depth, leveraging its 24 physical cores. For ResNet-18, the GNR platform achieves a peak 
speedup of \textbf{12.37$\times$} at 24 threads. To ensure maximum performance stability 
and mitigate the latencies associated with logical core sharing (Hyper-Threading), 
the GNR workload was explicitly pinned to physical cores 0--23 using \texttt{taskset}.

A notable ``performance cliff'' is observed when the workload extends beyond 24 
threads (see Fig.~\ref{fig:speedup_modern}). As the thread count $t$ exceeds the 
number of pinned physical cores, the speedup drops sharply to approximately $5.6\times$ 
at $t=48$. This regression is attributed to CPU oversubscription, 
wherein the operating system is forced to multiplex 48 software threads across 
only 24 available physical cores. This resource contention leads to 
increased context-switching overhead and severe cache thrashing, as threads frequently evict each other's data from the L3 cache. 
These results confirm that for compute-intensive ResNet inference, performance 
is optimized by aligning the thread count with the physical core count within 
a pinned affinity mask.

\subsection{Analysis of Findings}
The findings presented in Figures~\ref{fig:full_throughput_comparison}, \ref{fig:full_latency_comparison}, 
and \ref{fig:full_performance_comparison} elucidate three consistent patterns that account for the differing 
scaling limitations of legacy versus modern CPU-only inference systems: (i) the advantages of batching are 
contingent upon cache residency and available bandwidth, (ii) the scalability of threads is constrained by 
the availability of physical core resources and the shared memory hierarchy, and (iii) tail latency remains 
predominantly deterministic until the system experiences oversubscription.

\begin{figure*}[t]
  \centering
  \begin{subfigure}[t]{0.48\textwidth}
    \centering
    \includegraphics[width=\linewidth]{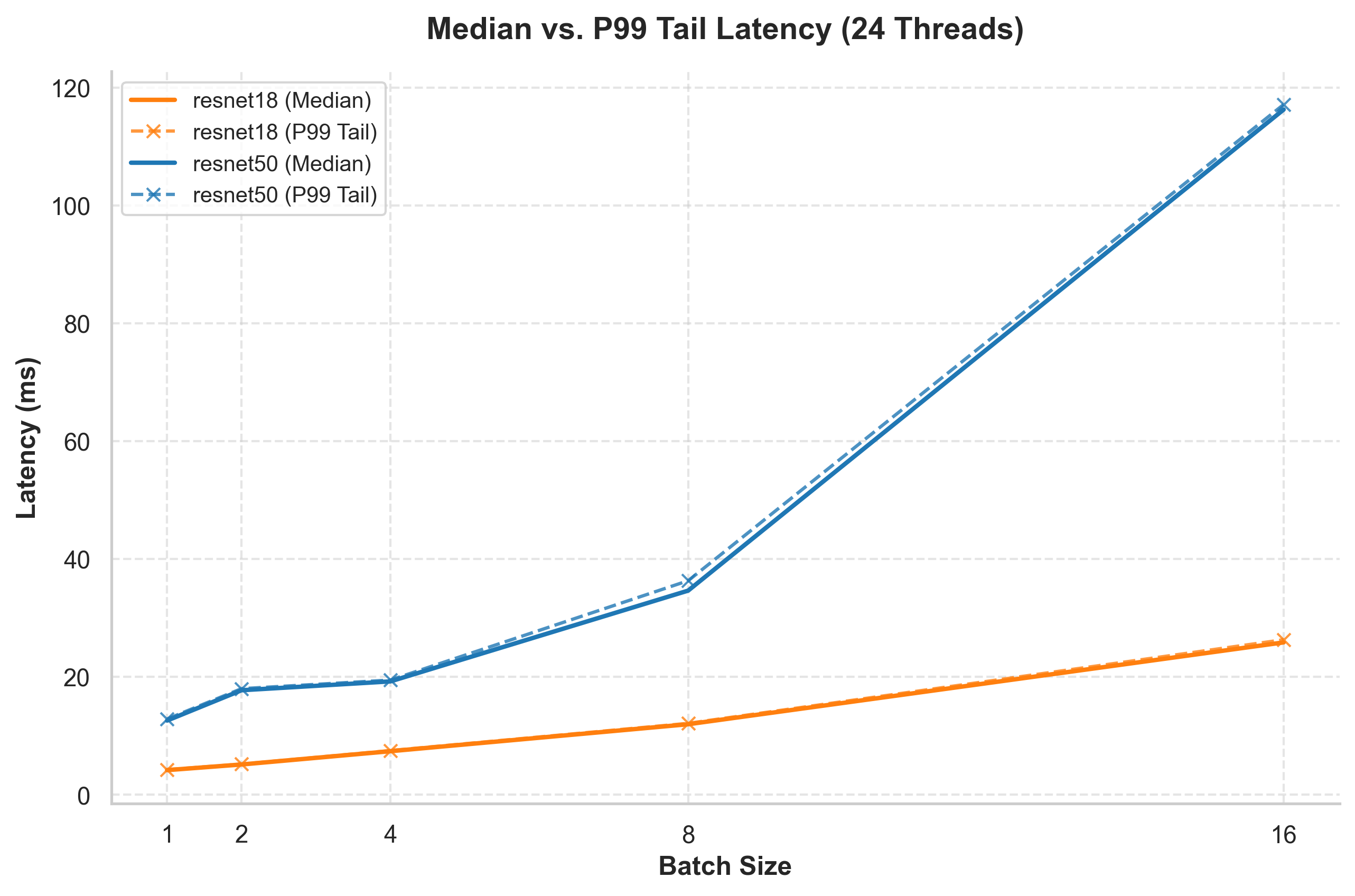}
     \label{fig:figure_tail_latency_24threads}
  \end{subfigure}
  \hfill
  \begin{subfigure}[t]{0.48\textwidth}
    \centering
    \includegraphics[width=\linewidth]{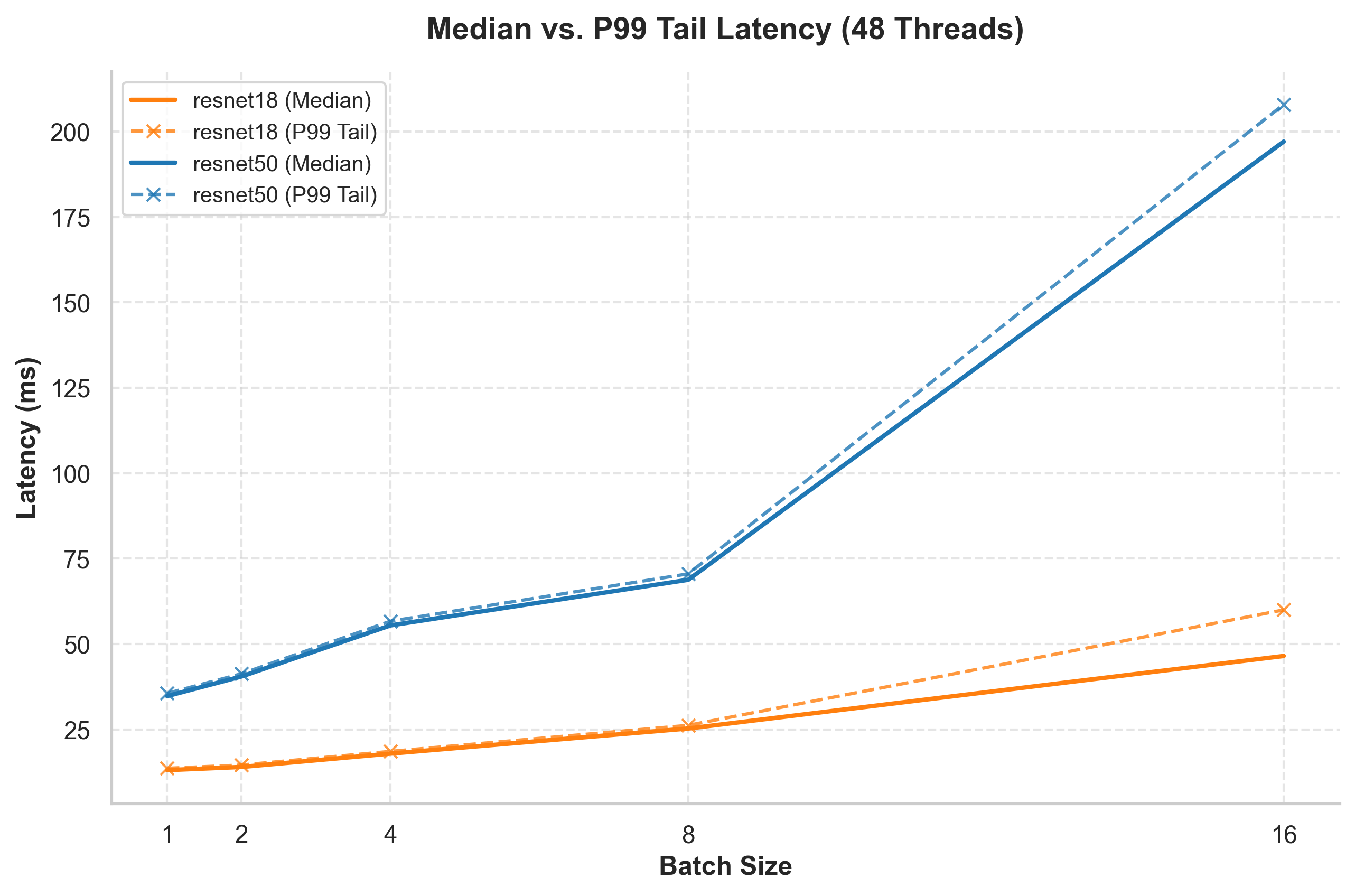}
     \label{fig:figure_tail_latency_48threads}
  \end{subfigure}

  \caption{Tail latency under physical-core saturation and oversubscription on the Granite Rapids platform.
At 24 threads, P99 closely tracks the median across batch sizes, indicating deterministic execution.
At 48 threads, P99 exhibits a modest divergence from the median, reflecting contention-induced variability
under oversubscription.
}
  \label{fig:tail_latency_oversubscription}
\end{figure*}

\textbf{1) Explanation for the Plateau in Legacy Throughput Beyond $B=4$.} On the legacy Xeon E5-2403 v2 processor, 
throughput experiences improvement from $B=1$ to $B=4$ due to the batching process, which effectively amortizes 
fixed per-inference overheads such as Python dispatch, operator setup, and framework scheduling. However, 
these gains rapidly diminish once the working set, comprising weights, activations, and intermediate feature 
maps, surpasses the cache hierarchy's efficient reuse capacity. With only four cores and a 10 MB shared L3 cache, 
the ResNet-50 weights and intermediate tensors cannot remain resident within the cache, necessitating frequent 
accesses to the main memory over DDR3. Beyond this point, additional batching results in increased memory traffic 
and cache pressure at a rate that outpaces the increase in useful compute utilization. This leads to the observed 
plateau: while latency increases with batch size, throughput does not improve proportionally. In practical 
terms, this indicates that the legacy platform reaches its effective saturation point early, rendering batching 
an ineffective strategy beyond small batch sizes.

\textbf{2) Why Granite Rapids scales to a higher batch ``sweet spot'' (peak near $B=8$).} The contemporary Granite 
Rapids-class system reconfigures the equilibrium between computation and memory in two significant ways. Firstly, 
it incorporates a substantially larger last-level cache and a high-bandwidth DDR5 memory subsystem, thereby 
enhancing the volume of reusable model state and activations accessible from faster memory tiers. Secondly, 
the architecture integrates AI-oriented execution support (e.g., AMX), which augments effective computational 
throughput on tensor-intensive kernels. Collectively, these enhancements extend the range where batching is advantageous: 
throughput continues to rise up to $B=8$ as the platform can maintain elevated arithmetic utilization before 
memory stalls become predominant. Beyond this threshold, throughput begins to decline (e.g., at $B=16$) due 
to the rapid expansion of the activation working-set size, rendering the memory subsystem and shared cache 
as limiting factors. Notably, the peak throughput of the modern platform is achieved at a batch size that still 
ensures practical median latency, which is the optimal condition for near-real-time inference.

\textbf{3) Thread scaling is effective up to physical cores, then collapses under oversubscription.} Thread 
sweeps indicate that increasing intra-operation threads yields significant performance improvements up to the 
number of physical cores, particularly on Granite Rapids. When the workload is confined to cores 0--23, the 
optimal operating point corresponds to $T=24$, as each software thread is mapped to an independent physical 
core. Beyond this threshold, increasing the number of threads does not generate additional computational resources; 
rather, it compels the operating system to time-slice more threads on the same cores. This results in context-switch 
overhead, heightened contention for shared resources (such as LLC capacity, memory controllers, and execution pipelines), 
and destabilizes per-iteration progress. As a result, a pronounced performance decline is observed: throughput 
decreases sharply while tail latency increases. Practically, this underscores that CPU-only inference should 
regard \emph{physical cores} (as opposed to logical threads) as the primary scaling limit, and that oversubscription 
is detrimental to compute-intensive CNN inference.

\textbf{4) Tail latency tracks the median until the system enters contention regimes.}
As shown in Figure~\ref{fig:tail_latency_oversubscription}, across most evaluated configurations on the
Granite Rapids platform, the 99th percentile latency (P99) closely aligns with the median,
indicating minimal tail amplification and suggesting that inference latency is largely governed by
deterministic architectural factors such as cache hierarchy behavior, memory bandwidth limits, and
predictable computational demand.
This behavior is particularly evident in parity experiments where platforms are evaluated at
equivalent thread counts.

\textbf{5) Operational takeaway: distinct optimization strategies by hardware tier.} The legacy platform should be 
regarded as a latency-constrained node, where batching beyond $B \approx 4$ yields minimal throughput benefits and 
significantly increases response time. For such systems, deploying numerous $B=1$ (or small-$B$) instances or utilizing 
lighter models is generally more effective than batching. In contrast, Granite Rapids-class CPUs accommodate a broader 
efficient operating range: batching up to $B \approx 8$ and threading to physical cores can achieve substantial throughput 
improvements while maintaining acceptable latency. Nonetheless, the performance decline beyond physical cores indicates 
that meticulous CPU affinity management and avoiding oversubscription are essential to sustain both throughput and 
tail latency stability.

\section{Implications for Batched Inference Workloads}
\label{sec:implications}

These results highlight a critical generational shift in CPU-based inference: while batching on 
legacy hardware is often a zero-sum trade-off between latency and utilization, modern architectures 
like Granite Rapids (GNR) can leverage batching to achieve significant throughput gains. On the legacy Xeon, 
additional batching beyond $B=4$ increases latency without improving throughput, representing an 
architectural mismatch. In contrast, the Granite Rapids platform achieves a \textbf{2.9$\times$ throughput 
improvement} via internal scaling ($B=1$ vs. $B=8$), demonstrating that modern CPUs can effectively 
scale inference workloads given sufficient cache and memory bandwidth.

To better understand the observed early saturation on legacy platforms compared to the deeper scaling of GNR, 
we examine the architectural factors at play.

The legacy Xeon E5 platform lacks AI-specific instruction sets such as Vector Neural Network Instructions (VNNI) 
or Advanced Matrix Extensions (AMX)~\cite{intel_vnni, intel_amx}. As a result, matrix operations are decomposed 
into sequences of general-purpose SIMD instructions, leading to high instruction counts and register pressure. 
Furthermore, the legacy 10~MB L3 cache and DDR3 memory are unable to sustain the data movement required for 
larger batch sizes. 

In contrast, the Granite Rapids server represents a specialized evolution. With a \textbf{144~MB Last-Level 
Cache (LLC)} and \textbf{DDR5-6400} memory, GNR can maintain larger activation tensors within high-speed memory 
tiers, significantly delaying the onset of memory-bound stalls. While the GNR platform still encounters a performance 
ceiling at high thread counts (particularly when oversubscribing logical cores), its baseline peak 
performance—reaching \textbf{230.9~IPS} for ResNet-50—makes it a viable alternative to dedicated accelerators 
for many edge and datacenter inference tasks.

These findings motivate the adoption of different strategies based on hardware vintage. For legacy deployments, 
throughput is best improved through model-level optimizations or horizontal scaling of $B=1$ instances. For modern 
platforms like Granite Rapids, significant gains can be realized through vertical scaling (batching) and the use 
of software toolchains such as Intel’s OpenVINO or TVM, which can leverage the platform's advanced vectorization 
and hardware-aware scheduling to further bridge the gap between CPUs and dedicated AI 
accelerators~\cite{openvino_intel, chen2018tvm}.

\begin{figure}[t]
    \centering
	\begin{subfigure}[b]{0.48\textwidth}
        \centering
        \includegraphics[width=\textwidth]{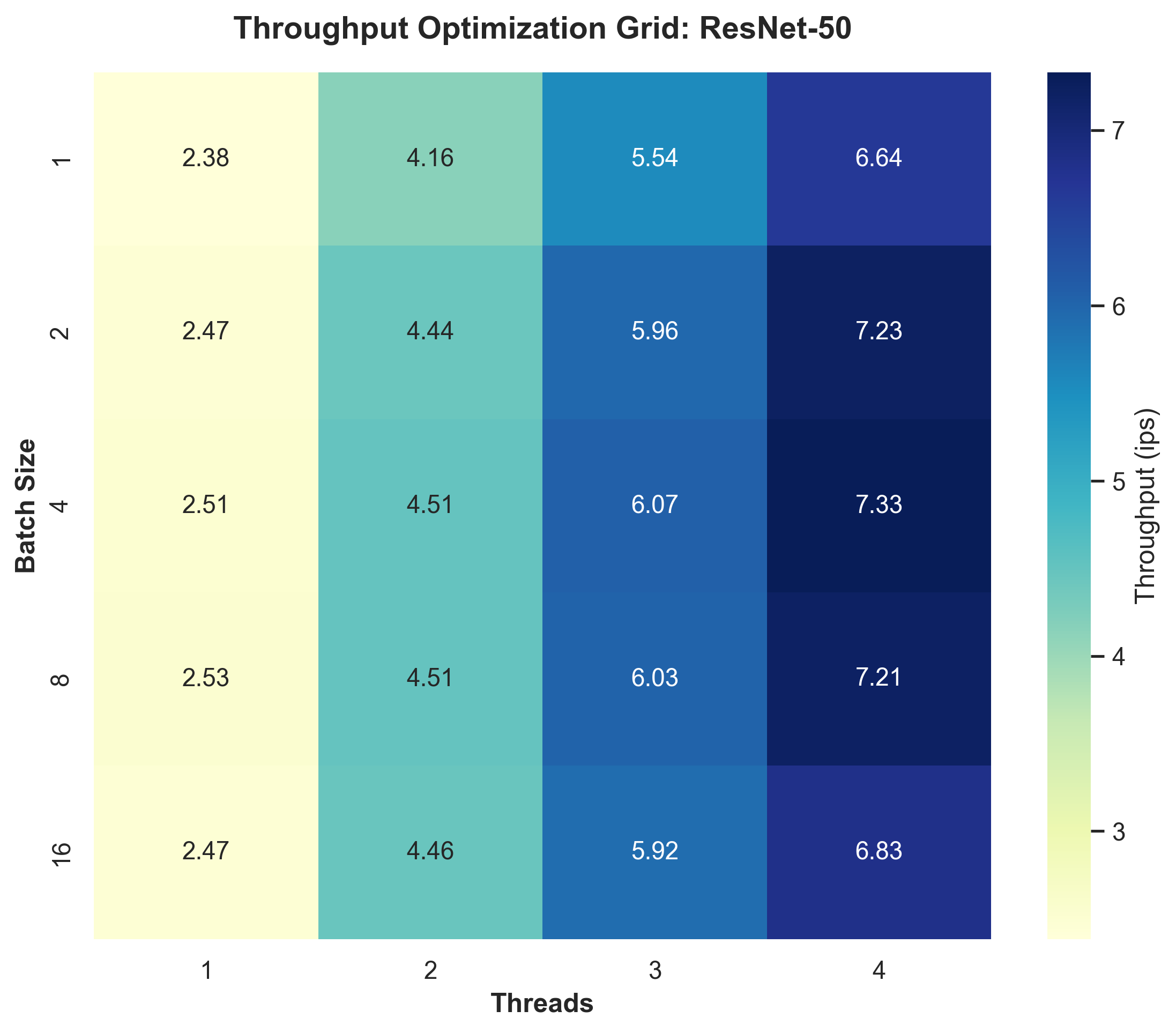}
        \caption{Legacy Xeon E5-2403 v2}
        \label{fig:heatmap_legacy}
    \end{subfigure}
    \hfill
    \begin{subfigure}[b]{0.48\textwidth}
        \centering
        \includegraphics[width=\textwidth]{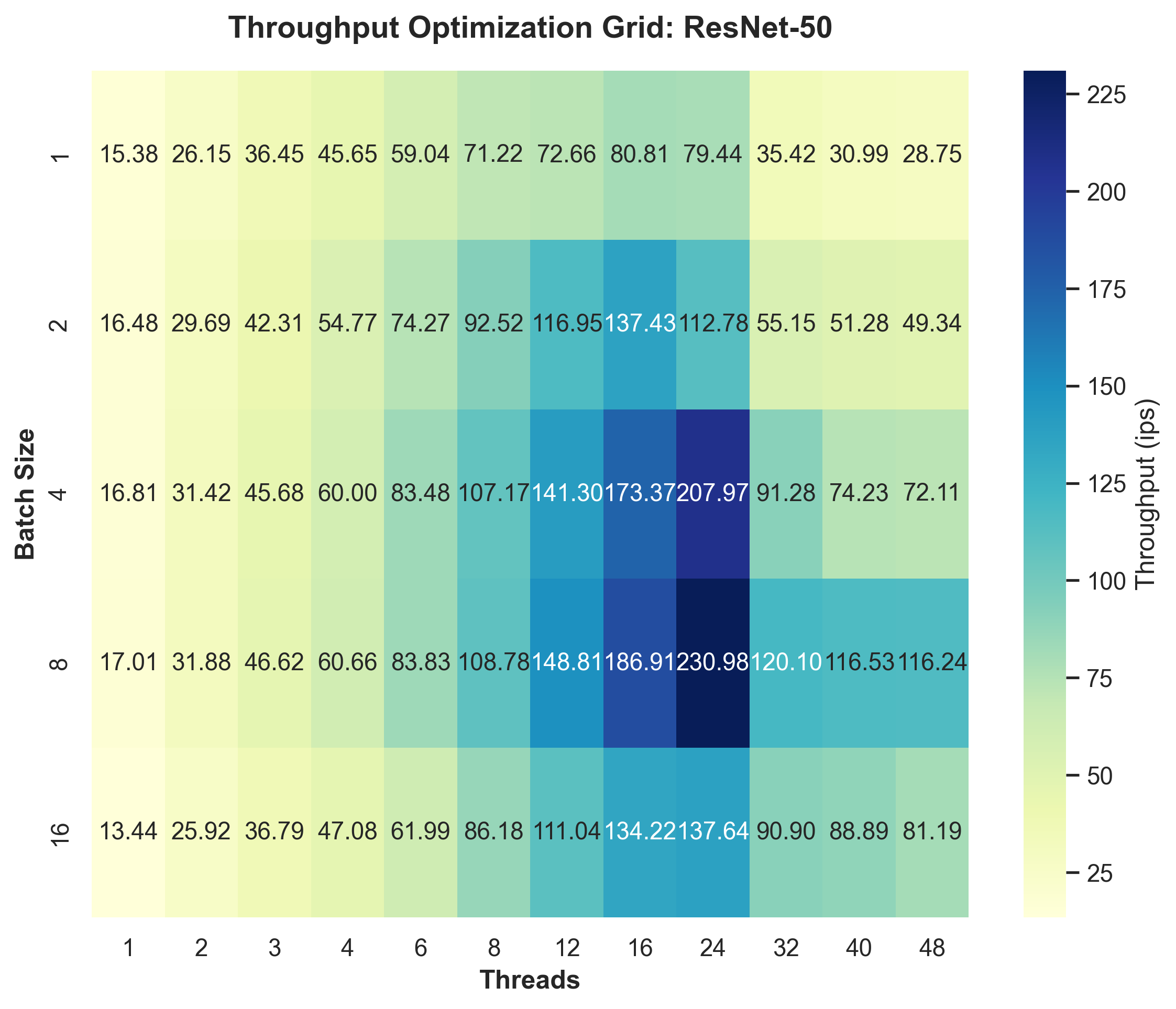}
        \caption{Modern Xeon Granite Rapids (GNR)}
        \label{fig:heatmap_gnr}
    \end{subfigure}
	
	\caption{Throughput optimization heatmaps for ResNet-50 inference across varying batch sizes and 
	thread counts. The comparison reveals a sharp contrast in architectural efficiency: (a) shows immediate saturation on the legacy platform due to limited vector resources and memory bandwidth, while (b) demonstrates the high-parallelism scaling and significantly expanded throughput envelope of the modern GNR architecture.}
    \label{fig:performance_heatmap_comparison}
\end{figure}

Figure \ref{fig:performance_heatmap_comparison} provides a localized view of the throughput landscape, 
measured in images per second (IPS), across the primary control variables: batch size ($B$) 
and thread count ($T$). The heatmaps for both platforms reveal a distinct "diminishing returns" 
gradient, though the scale and saturation points differ by an order of magnitude.

The visualization highlights three critical performance behaviors:

\begin{itemize}
    \item \textbf{Thread Dominance and Scaling Depth:} On the legacy Xeon, increasing $T$ from 
	1 to 4 yields a $2.8\times$ throughput gain (2.38 to 6.62~IPS). The Granite Rapids platform 
	demonstrates significantly deeper scaling, where increasing $T$ from 1 to 24 (the physical 
	core limit) at $B=1$ results in a \textbf{5.2$\times$ improvement} (15.26 to 79.35~IPS). 
	On both platforms, throughput is far more sensitive to multithreaded compute parallelism 
	than to batch-level amortization at low thread counts.
    
    \item \textbf{The Saturation Plateau and Performance Cliff:} A clear performance ceiling 
	is observed on both architectures. The legacy Xeon peaks at $T=4, B=4$ (7.33~IPS) before 
	plateauing. Granite Rapids reaches a much higher ceiling of \textbf{230.9~IPS} at $T=24, 
	B=8$. However, GNR exhibits a sharp ``performance cliff'' beyond 24 threads; because the 
	task was pinned to physical cores 0--23, increasing threads to 48 forces oversubscription, 
	causing throughput to collapse from 230.9~IPS back to 116.24~IPS. This visualizes the 
	point where context-switching overhead and execution unit contention override any architectural 
	benefits.

    \item \textbf{Optimal Operating Points:} The ideal configuration for the legacy Xeon is pinpointed 
	at $T=4, B=4$ (7.33~IPS). In contrast, the Granite Rapids server reaches its peak performance 
	with a setup of \textbf{$T=24, B=8$}, delivering 230.98~IPS.

	Beyond this point, throughput diminishes along two independent dimensions. 
	Firstly, oversubscribing the processor from $T=24$ to $T=48$ at $B=8$ 
	reduces throughput to \textbf{116.24~IPS}, a \textbf{49.7\% degradation}, 
	indicating significant logical-core contention and cache interference as 
	threads surpass the physical core density. Secondly, increasing the batch size from 
	$B=8$ to $B=16$ at a fixed $T=24$ decreases throughput to \textbf{137.64~IPS}, 
	a \textbf{40.4\% degradation}, which is attributed to an increased working-set size and memory 
	bandwidth pressure. When both factors are combined ($T=48$, $B=16$), 
	throughput further plummets to 81.19~IPS (\textbf{64.9\% degradation} relative to peak).

\end{itemize}

Our results demonstrate that modern CPU architectures like Granite Rapids provide a massive baseline 
performance boost—exceeding \textbf{30$\times$} the throughput of legacy systems—and offer a wider 
range of effective batch sizes. However, both platforms remain bound by deterministic architectural 
limits. While GNR can handle significantly larger workloads, the rapid saturation beyond physical core 
boundaries and large batch sizes provides empirical motivation for careful resource pinning and the 
use of optimized inference runtimes to stay within the hardware's efficient operating window.

\subsection{Roofline Analysis and Architectural Constraints}

We emphasize that the following Roofline-style analysis serves as an interpretive, back-of-the-envelope model. 
To understand the observed throughput saturation, we apply the Roofline analysis~\cite{williams2009roofline} 
to both legacy and modern platforms. The achievable performance is constrained by 
$P \le \min\left(P_{\max},\ \mathrm{OI} \cdot B_{\max}\right)$, with the ridge point defined as $\mathrm{OI}^\star = P_{\max}/B_{\max}$.

For the legacy Xeon E5-2403 v2, we estimate a theoretical peak compute throughput $P_{\max} 
\approx 115$~GFLOP/s and a sustained memory bandwidth $B_{\max} \approx 32$~GB/s, yielding a 
ridge point of $\mathrm{OI}^\star \approx 3.6$~FLOPs/byte. In contrast, the Granite Rapids (GNR) 
platform (Xeon 6 6521P) features 24 cores capable of AVX-512 vectorization. At a base frequency 
of 2.6~GHz, we estimate $P_{\max} \approx 4,000$~GFLOP/s (4~TFLOP/s). Supported by 12-channel 
DDR5-6400 memory with a sustained bandwidth $B_{\max} \approx 500$~GB/s, the GNR platform maintains 
a ridge point of:
\begin{equation}
\mathrm{OI}^\star_{\text{GNR}} \approx \frac{4000}{500} \approx 8.0\ \text{FLOPs/byte}.
\end{equation}

The shift in ridge point from 3.6 to 8.0 indicates that the modern platform requires significantly 
higher arithmetic intensity to reach its compute-bound regime. However, the most profound difference 
lies in the cache hierarchy. For ResNet-50 ($F \approx 3.8$~GFLOPs), model weights occupy 
$\approx 100$\,MB (FP32, framework-dependent layout). 

On the legacy platform, the 10\,MB L3 cache is an order of magnitude smaller than the model weights, 
forcing the system to stream weights from DDR3 memory for every inference. This keeps the effective 
operational intensity near the memory-bound slope. In contrast, the Granite Rapids server's 
\textbf{144\,MB Last-Level Cache (LLC)} exceeds the 100\,MB footprint of the ResNet-50 weights. 
This allows the model to become essentially ``cache-resident'' after the initial cold-start.

This architectural shift explains the scaling behavior observed in Figure~\ref{fig:throughput_legacy}. 
As the batch size $B$ increases, the legacy system transitions into a memory-bound regime when 
the effective data traffic exceeds
\begin{equation}
D \gtrsim \frac{3.8}{3.6} \approx 1.05~\text{GB per image}.
\end{equation}
Because GNR can cache weights, its effective data movement $D$ per image is drastically lower, 
consisting primarily of activation tensors. This allows GNR to sustain high throughput across a 
wider range of batch sizes before hitting the bandwidth wall. 

Our results confirm that the observed performance plateau on the legacy Xeon is a direct consequence 
of its low ridge point and insufficient cache capacity for modern CNNs. On the Granite Rapids platform, 
the combination of high-bandwidth DDR5 and a massive LLC shifts the performance bottleneck from basic 
data movement to the saturation of physical execution units, as evidenced by the peak throughput of 
\textbf{230.9~IPS} observed when utilizing all 24 pinned physical cores.

\begin{table}[t]
\centering
\caption{Operational Intensity (OI) and Performance Regimes for ResNet-50: Legacy Xeon vs. Granite Rapids.}
\label{tab:oi_resnet50_comp}
\small
\begin{tabularx}{\columnwidth}{l c X X}
\hline
\textbf{Data Moved ($D$)} & \textbf{OI} & \textbf{Legacy Regime} & \textbf{GNR Regime} \\ \hline
0.10~GB (Weights only)    & 38.0        & Compute-bound          & \textbf{Compute-bound} \\
0.47~GB (Ridge GNR)       & 8.0         & Compute-bound          & \textbf{Ridge Point} \\
1.00~GB (Activations)     & 3.8         & \textit{Near Ridge}    & Memory-bound \\
2.00~GB (Ltd. reuse)      & 1.9         & Memory-bound           & Memory-bound \\
4.00~GB (Heavy)           & 0.95        & Memory-bound           & Memory-bound \\ \hline
\multicolumn{4}{p{\dimexpr\columnwidth-2\tabcolsep}}{\scriptsize \textbf{Legacy $\mathrm{OI}^\star \approx 3.6$}: 115~GFL/s peak, 32~GB/s sustained DDR3} \\
\multicolumn{4}{p{\dimexpr\columnwidth-2\tabcolsep}}{\scriptsize \textbf{GNR $\mathrm{OI}^\star \approx 8.0$}: 4,000~GFL/s peak, 500~GB/s sustained DDR5 (upper-bound estimate)} \\
\multicolumn{4}{p{\dimexpr\columnwidth-2\tabcolsep}}{\scriptsize Note: GNR's 144\,MB cache allows Weights-only movement ($D=0.10$), maintaining compute-bound status.} \\ \hline
\multicolumn{4}{p{\dimexpr\columnwidth-2\tabcolsep}}{\scriptsize
\textbf{Note:} Data-movement values $D$ are illustrative scenarios for roofline regime classification (not directly measured).}
\end{tabularx}
\end{table}

\section{Reproducibility}

All experiments in this study were conducted using a deterministic benchmarking methodology. 
To maintain consistency across the decade-spanning hardware gap, both the legacy and modern platforms 
utilized a unified software stack consisting of \textbf{Python 3.6.8} and \textbf{PyTorch 1.10.1+cpu}. 

For the legacy Xeon E5-2403 v2 system, benchmarks were run on CentOS 7.5 using the default process 
scheduler across its 4 physical cores. For the Granite Rapids (GNR) Xeon 6521P platform, experiments 
were conducted on Ubuntu 20.04 Linux container (LXC) environment. To isolate physical core performance and eliminate the non-deterministic 
latency overhead of Hyper-threading, the GNR process was explicitly bound to physical cores 0--23 using 
the \texttt{taskset -c 0-23} utility. 

Measurement reliability was ensured by performing \textbf{100 iterations} for every batch size and thread 
count configuration, preceded by \textbf{20 warmup iterations} to allow for cache warming and frequency 
scaling stabilization. To mitigate the impact of system-level outliers on legacy hardware, 
the reported metrics represent the \textbf{median latency} and P99 tail latency across these repetitions. 

The complete benchmarking suite, including data collection scripts and raw results for both platforms,
will be made available in a public repository to support reproducibility and further analysis.
This baseline data serves as the foundation for our forthcoming comparative study between high-performance CPUs and modern 
GPU inference accelerators.

\section{Conclusion}

This study establishes a comparative baseline of CPU-only inference scaling, covering a decade 
of architectural evolution from a legacy Intel Xeon E5-2403 v2 to the modern Granite Rapids (GNR) 
Xeon 6521P platform. By systematically varying batch sizes and thread counts, we identified the 
practical performance limits and scaling bottlenecks of CPU-based inference under controlled, pinned, 
and reproducible conditions.

The findings indicate a notable generational performance gap: the Granite Rapids platform achieves up to 
approximately \textbf{32$\times$} higher throughput than the legacy system for ResNet-50. However, both platforms 
display deterministic saturation points. On the legacy system, throughput "flat-lines" almost immediately 
due to cache and bandwidth limitations. In contrast, the GNR platform's performance scales robustly up to the 
24-physical-core limit, but experiences a sharp decline beyond this point due to resource contention. 
This behavior underscores that while modern CPUs have significantly elevated the performance baseline, they 
remain constrained by the physical core count and the inherent memory-latency trade-offs of the CPU architecture.

By setting these clear and reproducible benchmarks, this study offers a concrete reference point for 
datacenter operators and researchers. We demonstrate that while Granite Rapids makes high-batch inference 
feasible for near-real-time workloads—achieving over \textbf{230~IPS}—there is still a distinct limit where 
general-purpose CPUs can no longer compete with the arithmetic density of specialized silicon. 

This work establishes the crucial foundation for a forthcoming evaluation, in which the Granite Rapids
results presented here will serve as a high-performance CPU benchmark against which accelerator-based
inference platforms can be compared. This progression will enable identification of the precise
crossover points at which transitioning from high-end CPU inference to specialized accelerated
computing becomes essential for scalable, production-grade AI environments.


\section{Future Work}

This study establishes a definitive baseline for CPU-only inference by characterizing the scaling limits, 
saturation behavior, and latency trade-offs of general-purpose processors under modern deep learning workloads. 
By quantifying where and why CPU-based inference reaches its practical limits, this work defines a clear reference 
point for evaluating alternative compute architectures.

Future work will extend this CPU-only baseline to GPU-accelerated inference in order to directly address the
limitations identified in this study. Using the same models, batch sizes, and experimental methodology, we will
evaluate inference-oriented GPU platforms to assess how massively parallel execution and high-bandwidth memory
alter throughput scaling, latency behavior, and tail stability relative to CPU-only execution. We will further
extend this analysis to larger model configurations to characterize the practical limits of GPU inference as
model size and working-set requirements increase. This approach enables a controlled architectural comparison
while preserving continuity with the CPU baseline established here.

In addition, we plan to investigate the impact of numerical precision (e.g., FP32 versus INT8) and hardware-supported 
tensor execution on GPU inference efficiency. These experiments will help isolate the roles of parallelism, memory 
hierarchy, and reduced-precision arithmetic in improving inference performance beyond what is achievable on general-purpose CPUs.

Collectively, this future work will establish the crossover point at which GPU acceleration transitions from a 
performance optimization to an operational necessity, providing actionable guidance for deploying scalable, 
production-grade inference systems.

\end{document}